\documentclass[fleqn,usenatbib]{mnras}

\usepackage{newtxtext,newtxmath}

\usepackage[T1]{fontenc}
\usepackage{adjustbox}

\DeclareRobustCommand{\VAN}[3]{#2}
\let\VANthebibliography\thebibliography
\def\thebibliography{\DeclareRobustCommand{\VAN}[3]{##3}\VANthebibliography}

\usepackage{graphicx}	

\usepackage{amsmath}	
\usepackage{amssymb}	
\usepackage{url}

\usepackage{pdflscape}
\usepackage{orcidlink}

\title[UV survey of mCP stars in Kepler field]{{\em GALEX} UV survey of magnetic chemically peculiar stars in the {\em Kepler} prime field}

\author[E. Bertone et al.]
{Emanuele Bertone,$^{1}$\thanks{E-mail: ebertone@inaoep.mx}
Stefan H\"ummerich,$^{2,3}$
Miguel Ch\'avez,$^{1}$\thanks{E-mail:mchavez@inaoep.mx}
Ernst Paunzen,$^{4}$
Klaus Bernhard,$^{2,3}$\newauthor
Daniel Olmedo,$^{1}$
Manuel Olmedo,$^{1,5}$
Eric. E. Mamajek,$^{6,7}$
Osvan M. Portilla-Narvaez,$^{1}$ \newauthor
Joshua Lara Sabala,$^{1}$
Alan Silva Castro$^{1}$
\\
$^{1}$Instituto Nacional de Astrof{\'\i}sica, \'Optica y Electr\'onica, Luis Enrique Erro 1, CP 72840, Tonantzintla, Puebla, Mexico\\
$^{2}$Bundesdeutsche Arbeitsgemeinschaft f{\"u}r Ver{\"a}nderliche Sterne e.V. (BAV), Munsterdamm 90, 12169 Berlin, Germany\\
$^{3}$American Association of Variable Star Observers (AAVSO), 49 Bay State Rd, Cambridge, MA 02138, USA\\
$^{4}$Faculty of Science, Masaryk University, Department of Theoretical Physics and Astrophysics, Kotl\'{a}\v{r}sk\'{a} 2, 611\,37 Brno,  Czechia\\
$^{5}$Universidad Aut\'onoma de Nuevo Le\'on, Pedro de Alba S/N, 66455 San Nicol\'s de los Garza, Nuevo Le\'on, Mexico\\
$^{6}$Jet Propulsion Laboratory, California Institute of Technology, 4800 Oak Grove Drive, Pasadena CA 91109, USA \\
$^{7}$University of Rochester, Department of Physics \& Astronomy, Rochester, NY 14627-0171, USA}

\date{Accepted XXX. Received YYY; in original form ZZZ}

\pubyear{2024}

\begin{document}
\label{firstpage}
\pagerange{\pageref{firstpage}--\pageref{lastpage}}
\maketitle

\begin{abstract}
Magnetic chemically peculiar (mCP) stars are strongly magnetic upper main-sequence stars that exhibit light rotational variability due to an uneven surface distribution of certain peculiar elements, which may appear in phase at certain wavelengths and in antiphase to the flux at other wavelengths. We present a study of the properties of photometric variability of a sample of confirmed mCP stars (mostly Ap/CP2 stars), mCP star candidates, and several non-CP stars in the near ultraviolet and visible wavelength regions based on observational data from the {\em GALEX} and {\em Kepler} prime missions. Antiphase variations between the near ultraviolet and optical light curves are observed in the majority of mCP stars. We investigate the presence of a correlation of the variability amplitudes in both wavelength regions with effective temperature, surface gravity, and metallicity and calculate model atmospheres, spectral energy distributions and synthetic light curves to connect our findings to theoretical models. While the theoretical calculations show that, at fixed abundances, a clear correlation between the light curve amplitude ratios and effective temperature is expected, our sample does not show any correlation with the investigated properties. This may be due to the highly individualistic abundance patterns of our sample stars, which are the main contributors to the line blanketing in different wavelength bands.
\end{abstract}




\section{Introduction}
\label{sect:introduction}

Ap/CP2 stars are upper main-sequence objects characterised by peculiar atmospheric abundances of elements such as Si, Fe, Sr, Cr, Eu, and the rare-earth elements as compared to the solar composition \citep[e.g.][]{preston74,renson09,ghazaryan2018}. They exhibit strong, stable, and globally organised magnetic fields with strengths of up to several tens of kilogauss \citep{babcock47,landstreet82,auriere07} and show a non-uniform surface distribution of elements, which is associated with the characteristics of the magnetic field and results in the formation of spots of enhanced or depleted element abundance. In these `chemical spots', flux is redistributed through line and continuum blanketing \citep[e.g.][]{wolff71,lanz96,shulyak10,krticka13}. As a consequence of the magnetic field being inclined to the rotation axis \citep{stibbs50}, most CP2 stars exhibit light, spectral, and magnetic field strength variations over the rotation period. Photometrically variable CP2 stars are conventionally referred to as $\alpha^{2}$ Canum Venaticorum (ACV) variables \citep{GCVS}.

The He-rich and the He-weak/CP4 stars \citep[e.g.][]{preston74,renson09,ghazaryan19} are in many respects the hotter analogues of the CP2 stars. Except for one subgroup (the He-weak PGa stars), they also possess strong and stable magnetic fields and present an inhomogeneous chemical surface composition, which leads to the same kind of variability as observed in the CP2 stars. Together with the CP2 stars, the He-peculiar stars are generally referred to by convention as magnetic chemically peculiar (mCP) stars.

The observed amplitude and shape of the light curves of ACV variables depend on the investigated wavelength regions. The light variations may appear in phase in different photometric passbands or vary in antiphase to the flux at other wavelengths \citep[e.g.][]{manfroid86,shulyak10}. 
A limited number of ACV stars have been studied at space-ultraviolet (UV) wavelengths. Antiphase variations of the flux between the far-UV and the optical wavelength regions is a common property, as it has been found in  CU Vir \citep{sokolov2000, krticka2019},  $\alpha^{2}$ CVn \citep{molnar73,sokolov2011}, a Cen \citep{sokolov2012},  $\iota$ Cas \citep{molnar1976}, HD~215441 \citep{leckrone1974}, $\phi$ Dra \citep{jamar1977,prvak2015}, $\theta$ Aur \citep{krticka2015}, and $\epsilon$ UMa \citep{molnar75}. It is due to a global flux redistribution to longer wavelengths, caused by the phase-dependent absorption. However, the pattern of this redistribution can be different from star to star. 
The wavelength interval where the flux remains almost unchanged over
the rotational cycle is called the `null-wavelength region'
\citep{molnar75} and may be located at different wavelengths. Therefore, the antiphase variations may not only occur between the far-UV and optical wavelengths.
In particular, the near-UV interval shows non-unique behaviour: its flux varies in phase with the visible in 
CU Vir \citep{sokolov2000, krticka2019}, 
a Cen \citep{sokolov2012, krticka2020},
$\phi$ Dra \citep{jamar1977,prvak2015}, and $\theta$ Aur \citep{krticka2015}, while it shows antiphase variation in $\alpha^{2}$ CVn \citep{molnar73,sokolov2011} and $\iota$ Cas \citep{molnar1976}. In the case of HD~215441, \citet{leckrone1974} finds a null wavelength at $\sim$2460~\AA, but does not present other NUV intervals, while $\epsilon$ UMa has not been studied in the space NUV interval.

Antiphase variations may sometimes also be observed in the optical region, where, in general, the light changes are in phase in the different photometric passbands. 
For example, antiphase variation of the $B$ and $V$ light curves of the CP2 star HD~240121 have been reported by \citet{groebel17}. \citet{faltova21} reported antiphase variability in the Zwicky Transient Facility \citep{ZTF1,ZTF2} $g$ and $r$ filters and employed this unique characteristic for the search for new CP2 star candidates. In general, ACV variables show an amazing diversity of light curve patterns, whose characteristics depend on the surface distribution of spots and the elements involved \citep{mikulasek2007}.

We here present a study of the properties of the photometric variability of a sample of mCP stars (mostly CP2 stars), mCP star candidates and several non-CP stars in the ultraviolet (UV) and visible regions based on observational data of the space telescope GALEX \citep{martin2003,bianchi2017} and the {\em Kepler} prime mission \citep{borucki2010}. We investigate the presence of a correlation of the variability amplitudes with stellar parameters that affect the intensity of the absorption, in particular effective temperature, and calculate synthetic light curves, model atmospheres, and spectral energy distribution (SED) profiles of diverse chemical compositions to connect our findings to theoretical models.

Data sets are described in Sect.~\ref{sec:observations}. The objects of our study, the sample of mCP stars in the {\em Kepler} prime field, are discussed in Sect.~\ref{sec:cpsample}. Next, we illustrate the procedure used to determine the near-ultraviolet (NUV) variability amplitude and its phase difference to the light changes observed in the visible region (Sect.~\ref{sec:analysis}). Results are presented in Sect.~\ref{sec:results}, notes on some individual objects in Sect.~\ref{sec:notes} and final conclusions are exposed in Sect.~\ref{sec:conclusions}.

\section{Observational data bases}
\label{sec:observations}

\subsection{The GALEX NUV observations}

The prime field of the {\em Kepler} mission \citep{borucki2010} has rapidly become one of the most observed and studied stellar fields. It is located in the northern hemisphere and is centred at R.A. = 19h22m40s and Dec =$+44^{\circ}30'00{\arcsec}$ (J2000), not far from Vega. Its more than 100 square degrees have been observed by the {\em GALEX} space telescope during the Complete All-Sky UV Survey Extension (CAUSE), as part of a programme funded by the Cornell University (P.I. James Lloyd). The observations have been carried out with the NUV detector \citep{martin2003,siegmund2004} only (the FUV detector stopped working in May 2009), during 47 days in August-September 2012, a period that coincides with Quarter 14 of the {\em Kepler} observations.

More details on the {\em GALEX} CAUSE survey of the {\em Kepler} prime field are provided by \citet{olmedo2015}. This article also presents the {\em GALEX}-CAUSE {\em Kepler} (GCK) catalogue\footnote{Also available at \url{http://vizier.u-strasbg.fr/viz-bin/VizieR-3?-source=J/ApJ/813/100}} of about 660,000 NUV point sources, down to a limiting magnitude of NUV$\simeq$22.6~mag. Most GCK sources are included in the {\em Kepler} Input Catalog \citep[KIC;][]{brown2011}.

Due to the modality used by {\em GALEX} in the observations of the {\em Kepler} field
(multiple scans along a larger circle), each GCK source has been observed several times. The {\em GALEX} processing pipeline divided the whole area into 180 tiles: each tile was visited 17 times, on average, with a maximum of 22, for a total of 3,251 images. Therefore, besides measuring the brightness of the sources on the co-added images as it was done by \citet{olmedo2015}, it is also possible to extract their light curves, by separately reducing  each tile and perform the photometry of all detected sources.
This has already been done by our group \citep{bertone2020} and the catalogue of all NUV light curves of point sources in the {\em Kepler} field will be presented in a future publication (Olmedo et al., in preparation).
Data reduction has followed the same procedure as in \citet{olmedo2015}, with the only difference of using the background image of the {\em GALEX} pipeline\footnote{\url{http://galex.stsci.edu/GR6/?page=scanmode}}, rather than computing our own. However, in the NUV band, the detector background is negligible and the sky background level is very low: it amounts to a count rate of $10^{-3}$~cts$\,$sec$^{-1}$arcsec$^{-2}$ \citep{morrissey2007}\footnote{See also the {\em GALEX} technical documentation at \url{http://www.galex.caltech.edu/researcher/faq.html} and \url{https://asd.gsfc.nasa.gov/archive/galex/FAQ/counts_background.html}}; therefore, for a source of NUV=20~mag (AB system) and FWHM=5$\arcsec$, it would correspond to $\sim 1\%$ of the total flux.

\subsection{The Kepler light curves}

The {\em Kepler} spacecraft used a differential photometer with a 115 square-degrees field of view and an aperture of 0.95~m \citep{borucki2010}. The detectors consisted of 21 modules each equipped with two 2200x1024 pixel CCDs. The {\em Kepler} telescope produced single-passband light curves in the visible range ($\sim$4200--9000~\AA) with two different integration times in the long-cadence (LC), 29.4~min-sampling mode \citep{LCdata} and the short-cadence (SC), $\sim$1~min-sampling mode \citep{SCdata}. 

The prime {\em Kepler} mission's main goal was the detection of transiting planets to determine the frequency of Earth-like planets in or near the habitable zone of Sun-like stars \citep{Borucki08,borucki2010}. It was in operation for four years (2009 May 2 to 2013 May 8), until the loss of a second reaction wheel on the spacecraft. After that, {\em Kepler} entered a redefined mission called K2 \citep{howell14} that lasted another four years until the satellite's official retirement in October 2018.

{\em Kepler} produced long, quasi-uninterrupted and high-quality time series data. To optimise solar panel efficiency, the spacecraft completed a 90 degree-roll every three months; therefore, {\em Kepler} data are divided into four quarters each year. The final Data Release 25 includes 197,096 stars \citep{mathur2017}, which were observed in some or all Quarters 1-17.

{\em Kepler} has been very successful in its main goal and discovered thousands of exoplanet candidates. It has also enabled stellar variability analyses with unprecedented detail. For more information on the {\em Kepler} spacecraft, we refer to \citet{borucki2010} and \citet{koch10}.

\section{Sample selection, classification and light curve data}
\label{sec:cpsample}

We initially collected a sample of mCP stars and candidate mCP stars in and near the {\em Kepler} prime field that was chosen from the lists of \citet{hummerich2018} and \citet{Bauer-Fasching2024}. It was then supplemented with another set of mCP star candidates identified by S. H{\"u}mmerich (private communication).

This initial number of more than 100 objects decreased by applying the following selection criteria: 
(1) the stars must have been observed by both {\em GALEX} and {\em Kepler}; 
(2) the number of points in the {\em GALEX} light curve must be greater than three; 
(3) the star must be fainter than 14.7~mag in NUV (GCK catalogue).

We imposed the latter requisite as the photon counting {\em GALEX} detector suffers from loss of linearity at high count rates (\citealt{morrissey2007}; see also \citealt{olmedo2015}). This problem also affects the repeatability of the measurements and thus it may create a spurious variability.
With respect to criterion (1), we comment that some of the stars in the {\em Kepler} field observed by {\em GALEX} have no reliable photometric data due to the presence of a nearby very bright object or because of other instrumental artifacts.

Applying these criteria, we ended up with a final sample of 28 stars, of which 21 are present in the sample of \citet{hummerich2018}, who presented an investigation of the light variability of mCP stars using {\em Kepler} data. These authors identified candidate mCP stars via light-curve properties (in particular monoperiodic variability and light-curve stability) and used newly acquired and archival spectra to investigate these candidates.

The subsample of the 21 stars from \citet{hummerich2018} which entered our sample consists of 14 spectroscopically confirmed mCP stars (KIC~2853320, 3945892, 5739204, 5774743, 6715809, 6950556, 7628336, 8773445, 9541567, 10685175, 10905824, 10959320, 11154043, 11465134), three candidate mCP
stars (KIC~2969628, 3326428, 8362546), and four stars in which these authors could not establish chemical peculiarities (non-CP stars; KIC~5213466, 5727964, 8569986, 10082844). The non-CP stars have been selected as `control sample' and to check the spectroscopic results with the here described method.

The stars KIC~4171302, 5000179, 8386865, 9665384, 10090722, and 10096019 are from an unpublished list of mCP star candidates (H{\"u}mmerich, private communication). Two of these objects (KIC~8386865 and KIC~10090722) were subsequently confirmed as mCP stars by \citet{huemmerich20}.

Finally, the star KIC~7976845 was identified as an mCP star on the basis of its light variability properties \citep[e.g.][]{faltova21} and a significantly positive $\Delta a$ value  calculated from Gaia BP/RP low-resolution spectra \citep{paunzen05,carrasco21,paunzen22}. It is part of a recently published sample of mCP stars and candidates \citep{Bauer-Fasching2024}.

As described in the following section (Section \ref{sect:spect_class}), we use here our own spectroscopic observations and, where available, spectra from the Large Sky Area Multi-Object Fiber Spectroscopic Telescope (LAMOST) to investigate and confirm the mCP star candidates KIC~2969628, 3326428, 4171302, 5000179, 9665384, and 10096019 as bona-fide CP2 stars.

In summary, the final sample of stars that was investigated with GALEX UV and {\em Kepler} photometry in this study consists of 22 confirmed CP2 stars, one photometrically confirmed mCP star (KIC~7976845), one mCP star candidate (KIC~8362546), and four non-CP stars (KIC~5213466, 5727964, 8569986, 10082844).

Relevant parameters of our sample stars are reported in Table~\ref{tab:sample}. The first two columns contain the KIC and the GCK identifiers. Columns three and four report another identifier from the SIMBAD or VizieR databases \citep{SIMBAD} and the stellar spectral type (sources are identified in the table). Column five denotes the CP star classification (CP2 = CP2 star; cand = candidate mCP star; nonCP = non-CP star; photCP = mCP star identified by photometric criteria alone). J2000 coordinates from Gaia DR3 \citep{GAIA1,GAIA2,GAIA3} are provided in the next two columns. The rotational periods presented in column eight are based on literature values (as indicated in the notes to the table) and are presented here rounded to five decimal places. The last two columns contain the brightness in the {\em Kepler} and {\em GALEX} NUV passbands.

\begin{table*}
\caption{Relevant properties of the sample of mCP stars. Unless indicated otherwise in the notes to this table, spectral types and period values are from \citet{hummerich2018}.}
\label{tab:sample}
\begin{adjustbox}{max width=\textwidth}
\begin{tabular}{lllllcclcc}
\hline
KIC & GCK & Other id. & Sp.Type & Class. & R.A. & Dec. & Period & Kepler & NUV \\
    &     &           &         &   & (deg)& (deg)& (d)    & (mag)  & (mag) \\
\hline
 2853320 &  GCK\_J19263019+3802516 &  2MASS J19263017+3802518 &        A0 V Si & CP2         & 291.625721 & +38.047703 &   5.06533 &  13.70 &  15.67 \\
 2969628 &  GCK\_J19025476+3809570 &           TYC 3120-750-1 &    A7 V SrCr$^{(1)}$  & CP2 &   285.728088 & +38.165967 &   1.97361 &  11.70 &  15.02 \\
 3326428 &  GCK\_J19061861+3824207 &  2MASS J19061861+3824209 &  kB9:hA5 V SrCrEu$^{(1)}$  & CP2 & 286.577542 & +38.405769 &   7.70042 &  13.41 &  16.98 \\
 3945892 &  GCK\_J19160898+3900250 &   TYC 3121-127-1 &   A2 V SiSrCrEu &         CP2 & 289.037304 & +39.006961 &   4.08323 &  12.23 &  15.35 \\ 
 4171302 &  GCK\_J19390868+3916088 &           TYC 3135-491-1 &     kA3hA5 V SrCrEu$^{(2)}$      & CP2   & 294.785890 & +39.269001 &   8.73500$^{(a)}$ &  11.81 &  14.95 \\ 
 5000179 &  GCK\_J19132601+4010289 &                 KOI-6485 &    kB9hA3 V SiCrSrEu$^{(2)}$  & CP2        & 288.358280 & +40.174660 &   3.63400$^{(b)}$ &  13.76 &  16.39 \\ 
 5213466 &  GCK\_J19523121+4023597 &  2MASS J19523118+4023594 &           A1 V    & nonCP     & 298.129900 & +40.399825 &   2.81951 &  13.07 &  16.51 \\
 5727964 &  GCK\_J19501553+4058358 &  2MASS J19501553+4058357 &           A6 V   & nonCP      & 297.564708 & +40.976564 &   1.63014 &  12.93 &  16.41 \\
 5739204 &  GCK\_J19591601+4056163 &  2MASS J19591596+4056166 &       B9 V SiEu    & CP2    & 299.816529 & +40.937928 &   1.81123 &  13.61 &  17.11 \\
 5774743 &  GCK\_J19042026+4101441 &  TYC 3124-443-1 &       A3 V SiCr   & CP2      & 286.084513 & +41.029186 &   4.07357 &  12.14 &  15.41 \\
 6715809 &  GCK\_J19511202+4206261 &  2MASS J19511198+4206264 &     A1 V SiCrEu  & CP2      & 297.799937 & +42.107297 &   4.19793 &  12.50 &  16.05 \\
 6950556 &  GCK\_J19294384+4229306 &  2MASS J19294376+4229306 &         A0 V Si   & CP2     & 292.432367 & +42.491831 &   1.51179 &  12.75 &  15.02 \\
 7628336 &  GCK\_J19493626+4313081 & TYC 3148-183-1 &  A3 V SiSrCrEu     & CP2     & 297.401025 & +43.218967 &   2.53883 &  11.35 &  14.77 \\
 7976845 &  GCK\_J19484629+4343516 & UCAC4 669-077857 & n/a & photCP           & 297.192930 & +43.730991 &   1.83492$^{(c)}$ &  15.52 &  18.87 \\ 
 8362546 &  GCK\_J19224728+4419142 &  2MASS J19224722+4419143 & n/a & cand      & 290.696825 & +44.320619 &   1.10814 &  15.69 &  16.60 \\
 8386865 &  GCK\_J19534140+4419393 &  2MASS J19534139+4419399 &   A0 V CrEu$^{(3)}$   & CP2      & 298.422490 & +44.327751 &   1.25800$^{(d)}$ &  12.02 &  15.22 \\ 
 8569986 &  GCK\_J19420666+4438591 &  2MASS J19420663+4438592 &            A2 V   & nonCP     & 295.527642 & +44.649778 &   3.13318 &  13.43 &  16.36 \\
 8773445 &  GCK\_J19531423+4457123 &  2MASS J19531426+4457124 &  A0 IV SiCrSrEu  & CP2      & 298.309417 & +44.953425 &   3.66078 &  13.84 &  17.25 \\
 9541567 &  GCK\_J19493708+4607121 & KOI-7190 &     A9 V SrCrEu        & CP2 & 297.404237 & +46.119989 &   2.24569 &  11.87 &  14.74 \\
 9665384 &  GCK\_J19504933+4622500 & UCAC4 682-072593 & kA0hB9 III Si$^{(2)}$ & CP2 & 297.705410 & +46.380520 &   1.04615$^{(c)}$ &  14.58 &  16.03 \\ 
10082844 &  GCK\_J19391261+4701086 &  2MASS J19391258+4701085 &            A0 V   & nonCP     & 294.802400 & +47.019017 &   2.08338 &  13.69 &  15.96 \\
10090722 &  GCK\_J19492598+4702164 &         UCAC4 686-073039 & B9.5 II-III EuSi$^{(3)}$ & CP2 & 297.357979 & +47.038344 &   6.00962$^{(e)}$ &  13.00 &  15.31 \\ 
10096019 &  GCK\_J19552479+4704589 &         UCAC4 686-074397 &        kA3hA5 V SrCrEu$^{(2)}$  & CP2    & 298.853240 & +47.082790 &   6.87600$^{(a)}$ &  12.39 &  17.23 \\ 
10685175 &  GCK\_J19541720+4757500 &                 KOI-7362 &        A4 V Eu  & CP2       & 298.571600 & +47.963928 &   3.10199 &  12.07 &  15.03 \\
10905824 &  GCK\_J18544470+4820245 &  2MASS J18544461+4820247 &       A1 V SiCr  & CP2      & 283.685917 & +48.340231 &   2.71954 &  12.81 &  15.30 \\
10959320 &  GCK\_J18481850+4828538 &         UCAC4 693-063535 &     A0 V SiCrSr  & CP2      & 282.076729 & +48.481633 &   2.44558 &  13.19 &  15.51 \\
11154043 &  GCK\_J19543527+4847042 &                 KOI-7414 &       A0 V SiCr  & CP2      & 298.646833 & +48.784622 &   4.52984 &  11.99 &  15.06 \\
11465134 &  GCK\_J19453889+4922275 &           TYC 3565-508-1 &         A0 V Si   & CP2     & 296.411767 & +49.374414 &   1.48781 &  12.40 &  14.74 \\
\hline
\multicolumn{10}{l}{Notes on spectral types: $^{(1)}$ This study, own spectroscopic observation. $^{(2)}$ This study, LAMOST spectrum. $^{(3)}$ \citet{huemmerich20}.} \\
\multicolumn{10}{l}{Notes on period values: $^{(a)}$ \citet{nielsen13}. $^{(b)}$ \citet{gao16}. $^{(c)}$ \citet{Bauer-Fasching2024}. $^{(d)}$ \citet{conroy14}. $^{(e)}$ \citet{balona17}.}
\end{tabular}
\end{adjustbox}
\end{table*}

\subsection{Spectral classification} \label{sect:spect_class}

Spectral classification was performed using own spectroscopic observations and spectra from LAMOST.

We acquired observations of KIC~3326428 and KIC~2969628 in July and September 2023 at the 2.1-meter telescope of the Observatorio Astrof{\i\i}sico Guillermo Haro, located in Sonora, Mexico, with a B\"oller \& Chivens spectrograph, equipped with an e2v 42-40 2040$\times$2040~px CCD. The instrument setup consisted of a 600 line~mm$^{-1}$ grating, at a blaze angle of 8.3$^\circ$, and a 200 $\mu$m wide slit; this configuration provided a
wavelength coverage between about 3870 and 4800\,\AA, with a spectral resolution of 3\,\AA\ for KIC~3326428 and 2.3\,\AA\
for KIC~2969628, due to an improved focusing of the spectrograph.
Multiple images of each spectra were co-added to reach a signal-to-noise ratio of about 10 for KIC~3326428 (observed during cloudy weather), and about 30 for KIC~2969628. 
The spectra were reduced using \textsc{IRAF} \citep{tody1986,tody1993}, following the standard procedure for spectroscopic data: bias subtraction, flat-field correction, cosmic-ray removal, wavelength calibration (through an internal HeAr lamp) and flux calibration, using standard stars from the European Southern Observatory list \citep{hamuy1992}.

The LAMOST survey \citep{lamost1,lamost2} employs a reflecting Schmidt telescope located at the Xinglong Observatory in Beijing, China, with an effective aperture of 3.6$-$4.9\,m and a field of view of 5$\degr$. LAMOST can take up to 4000 spectra in a single exposure (resolving power R$\sim$1800; limiting magnitude r\,$\sim$\,19\,mag; wavelength coverage  3700\,\AA\ to 9000\,\AA) and is therefore particularly suited for large-scale spectral surveys. Data products are released to the public in consecutive data releases and can be accessed via the LAMOST spectral archive.\footnote{\url{http://www.lamost.org}} The present study used spectra released in LAMOST DR4 \citep{DR4}.

Spectral classification was performed in the framework of the refined MK classification system following the methodology outlined in \citet{gray87,gray89a,gray89b}, \citet{gray94} and \citet{gray09}. For a precise classification and to identify peculiarities, the blue-violet (3800$-$4600\,\AA) spectral region was compared visually to, and overlaid with, MK standard star spectra, which were taken from the \textit{libr18} collections available from R. O. Gray's MKCLASS website.\footnote{\url{http://www.appstate.edu/~grayro/mkclass/}}

mCP stars exhibit several peculiarities that need to be taken into account during the process of classification. They often show weak or otherwise peculiar \ion{Ca}{ii} K line profiles, weak \ion{Mg}{ii} 4481~\AA\ lines and are markedly deficient in He \citep{gray09,ghazaryan2018}. In addition, mCP stars generally show enhanced and peculiar metallic lines. Therefore, the hydrogen-line profiles are generally the most accurate indicators of the actual temperature of these objects \citep{gray09}. Where appropriate, spectral types based on the \ion{Ca}{ii} K line strength (the k-line type) and the hydrogen-line profile (the h-line type) \citep{osawa65} were determined. As the metallic lines of most mCP stars are so peculiar that they cannot be used for luminosity classification, luminosity types were based on the wings of the hydrogen lines \citep{gray09}.

\begin{figure*}
    \includegraphics[width=1.5\columnwidth]{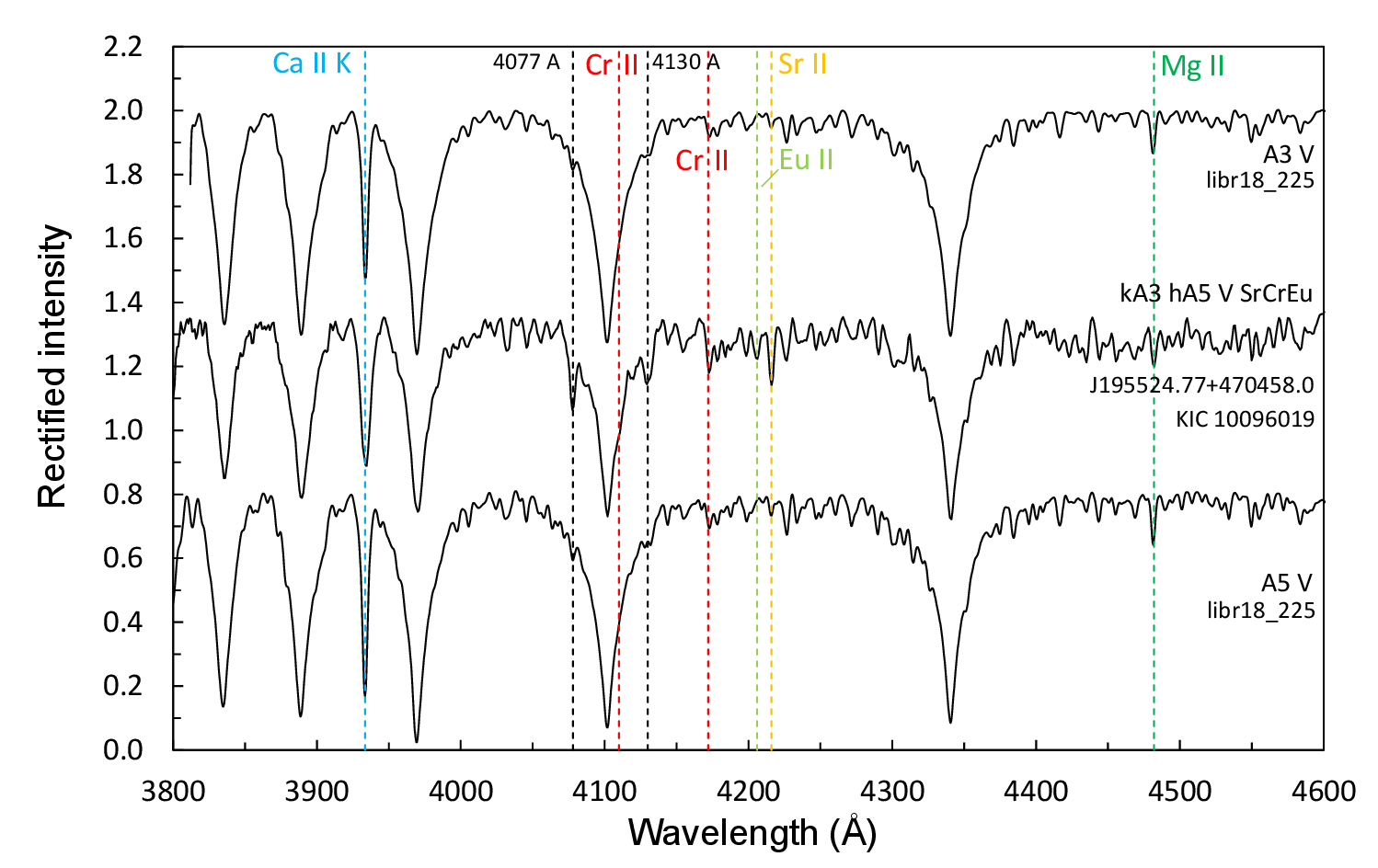}
    \caption{Blue-violet region of the LAMOST DR4 spectrum of the CP2 star KIC 10096019 = LAMOST J195524.77+470458.0 (middle spectrum), compared with two standard star spectra taken from the \textit{libr18\_225} collection. Some prominent lines and blends relevant to the classification of CP2 stars are identified. We note the peculiarly strong \ion{Cr}{ii}, \ion{Sr}{ii}, and \ion{Eu}{ii} features and the weak and unusually broad \ion{Ca}{ii} K line in the CP2 star.}
    \label{fig:showcase_spectra}
\end{figure*}

In this way, KIC~2969628 and 3326428 were confirmed as CP2 stars with own spectroscopic observations, while KIC~4171302, 5000179, 9665384, and 10096019 were confirmed as CP2 stars on the basis of LAMOST spectra. Figure \ref{fig:showcase_spectra} provides an example of this process. The final spectral types are included in Table \ref{tab:sample}.

\subsection{Light curves and phase diagrams in the NUV and the visible}

We searched and downloaded the {\em Kepler} light curves of our sample stars from the Mikulski Archive for Space Telescopes (MAST)\footnote{\url{https://archive.stsci.edu/kepler/data_search/search.php}}.
As the {\em GALEX}-CAUSE observations of the {\em Kepler} prime field have been carried out from 2012 August 3 to September 20, we extracted the {\em Kepler} data from the simultaneously acquired Quarter~14. We transformed the {\em Kepler} time system, expressed in barycentric Julian Date, which is used to correct for the motion of the spacecraft with respect to the center of mass of the Solar system, to the {\em GALEX} system, expressed in Julian Date (JD); we used the formula from \citet[][p.~17]{thompson2016}. We also defined the time $t$=0~d, which corresponds to JD=2456143.0, and selected the {\em Kepler} data points acquired in the time interval $0 \leqslant t \leqslant 47$~d because the {\em GALEX} data cover the interval $0.8 \lesssim t \lesssim 46.5$.
Two stars have not been observed by {\em Kepler} in Quarter 14: for KIC~4171302, we therefore extracted data from Quarter 13, while for KIC~9665384, we selected data from Quarter 9 ($\sim$1.2~yr before).
To produce the working light curve for each object, we first transformed the PDCSAP flux to magnitudes \citep[see][]{thompson2016} and we then subtracted the average magnitude, since we are only interested in analyzing the variability of the sources.
Each light curve typically has about 2230 data points. 

All {\em GALEX} light curves of our sample stars can be considered of good quality since all points have a signal-to-noise ratio SNR$>$10. Furthermore, we visually inspected a suitable region of the {\em GALEX} images of each visit to each star, in order to check whether the photometric measurements were affected by instrumental artifacts \citep{olmedo2015,bianchi2017}\footnote{We refer to the {\em GALEX} technical documentation for a more detailed explanation of the instrumental artifacts; see, e.g., \url{http://www.galex.caltech.edu/DATA/gr1_docs/GR1_Pipeline_and_advanced_data_description_v2.htm} and 
\url{http://www.galex.caltech.edu/wiki/Public:Documentation/Chapter_8\#Artifact_Flags}}. This analysis did not show any anomalies. For these light curves to be comparable with the {\em Kepler} ones, we subtracted the average NUV magnitude. These light curves have between 11 and 22 data points, with a median of 18 points.

We then folded the data of the light curves into phase diagrams by assigning phase $\phi$=0 to time $t$=0~d for all objects, assuming (as has been well studied) that the variability periods coincide with the rotation periods (see Table~\ref{tab:sample}). This implies that $\phi$=0 does not necessarily coincide with a specific feature of the light curve, such as the maximum or minimum of the brightness.

\begin{figure*}
\centering
\begin{tabular}{ccc}
\resizebox{54mm}{!}{\includegraphics{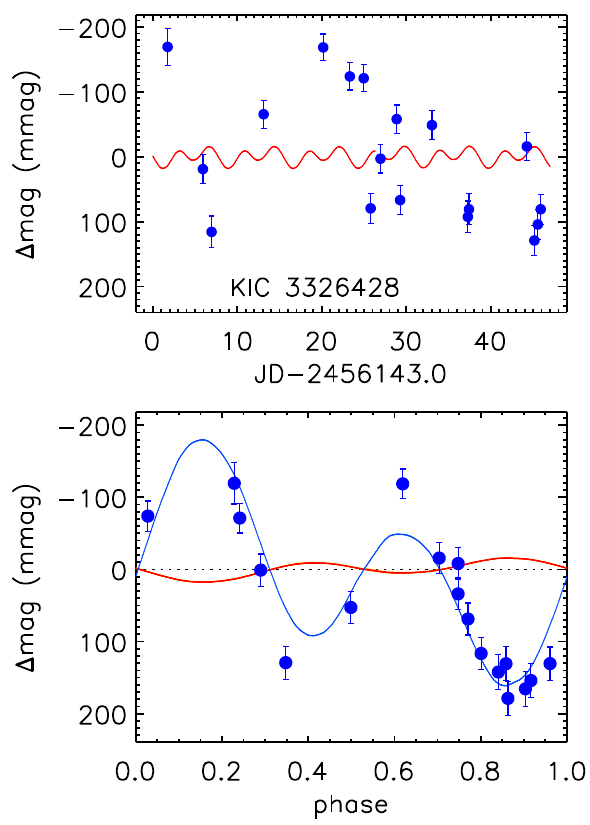}} &
\resizebox{54mm}{!}{\includegraphics{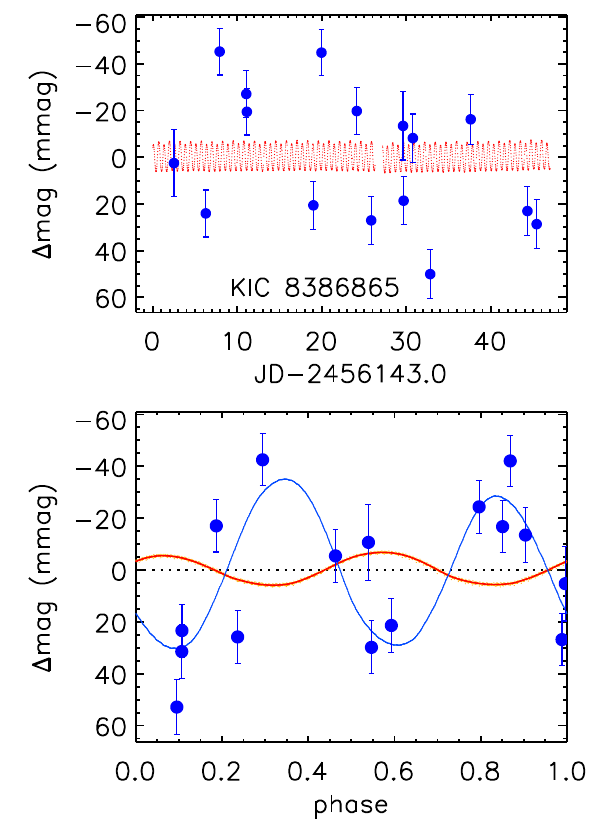}} &
\resizebox{54mm}{!}{\includegraphics{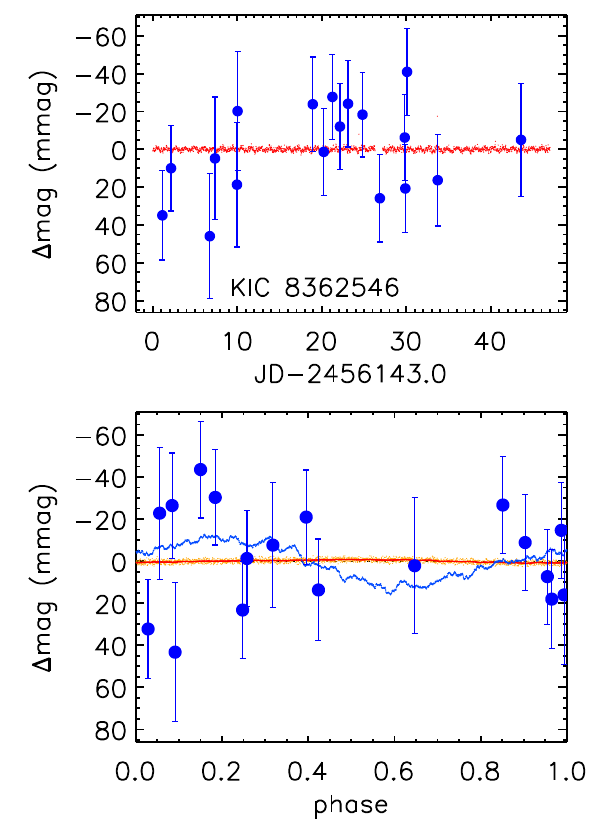}} 
\end{tabular}
\caption{ {\bf Upper panels}: {\em Kepler} (red dots) and {\em GALEX} (blue dots with error bars) light curves of the confirmed CP2 stars KIC~3326428 (kB9:hA5 V SrCrEu) and KIC~8386865 (A0 V CrEu), and the candidate CP2 star KIC~8362546. {\bf Lower panels}: phase diagrams for the same stars as in the upper panels using the periods listed in Table \ref{tab:sample}. The (barely visible) orange dots represent $\sigma$-clipped {\em Kepler} data; the smoothed curve is shown in red. The blue curve is the best fit of the {\em GALEX} measurements (blue dots with error bars), obtained from the genetic algorithm, as explained in Sect.~\ref{sec:aga}. Note that the {\em GALEX} points are shifted here by the offset values reported in Table~\ref{tab:agaresults}.} 
\label{fig:phase}
\end{figure*}

Some {\em Kepler} phase curves show outliers or instrumental drifts or a non-negligible dispersion in the brightness at the same phase. For the later analysis, we therefore had to produce an average curve. Hence, we first proceeded with an iterative 3$\sigma$ clipping and then computed an average value by smoothing the data with a moving boxcar of 50 points. As examples, in Fig.~\ref{fig:phase}, we show several light curves and the corresponding phase diagrams. 

\section{Assessing the NUV amplitude and phase difference}
\label{sec:analysis}

The number of the NUV points in the phase curves is too small for directly assessing the variability properties, such as the period and the amplitude. Therefore, we made the assumption that the {\em GALEX} NUV light curve has the same period and shape as the visible one, which has been shown to hold true for most ACV variables \citep[e.g.][cf. also Section \ref{sect:introduction}]{molnar73,molnar75}, and used two different methods to determine the amplitude of the NUV variation and the phase shift between the NUV and the visible curves. The first method makes use of an Asexual Genetic Algorithm \citep[AGA;][]{canto2009}, while in the second we fit the phase curves with harmonic polynomials, following the model of \citet{mikulasek2007}.

\begin{figure*}[!t]
\centering
\begin{tabular}{ccc}
\resizebox{56mm}{!}{\includegraphics{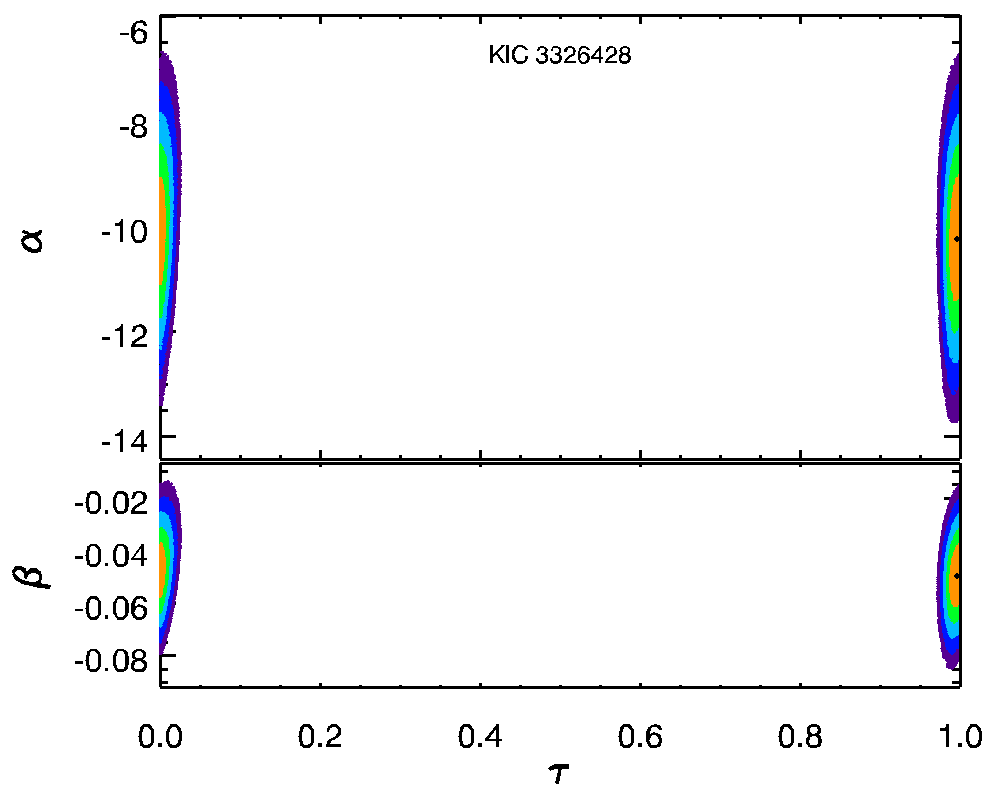}} &
\resizebox{56mm}{!}{\includegraphics{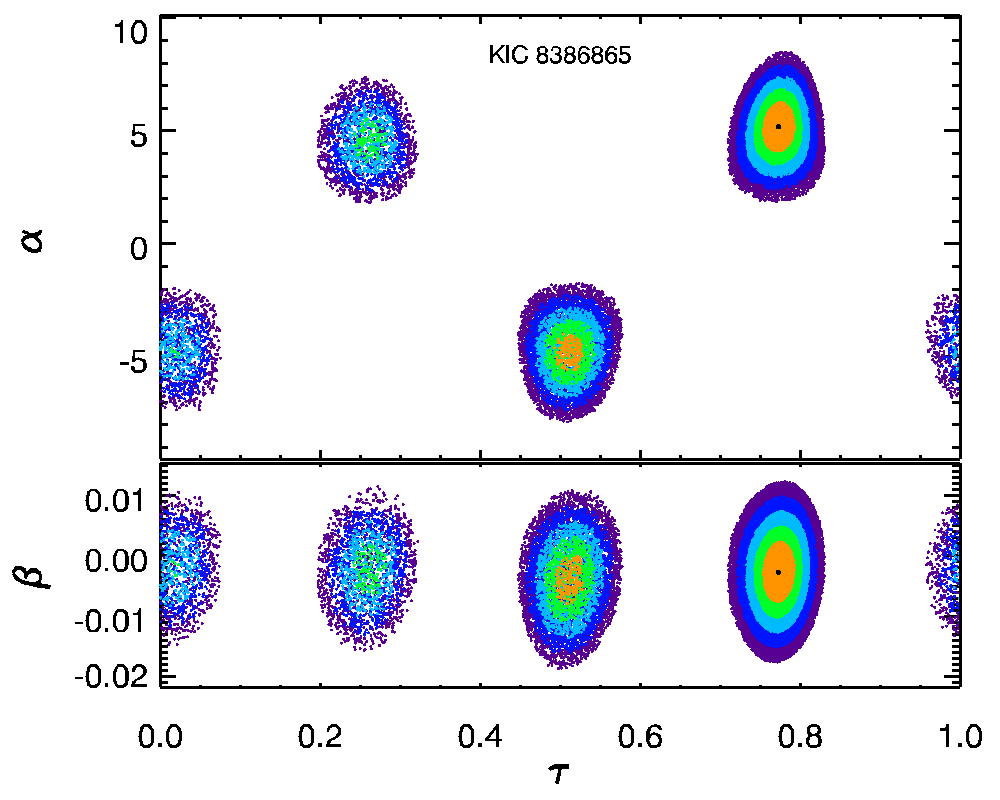}} &
\resizebox{56mm}{!}{\includegraphics{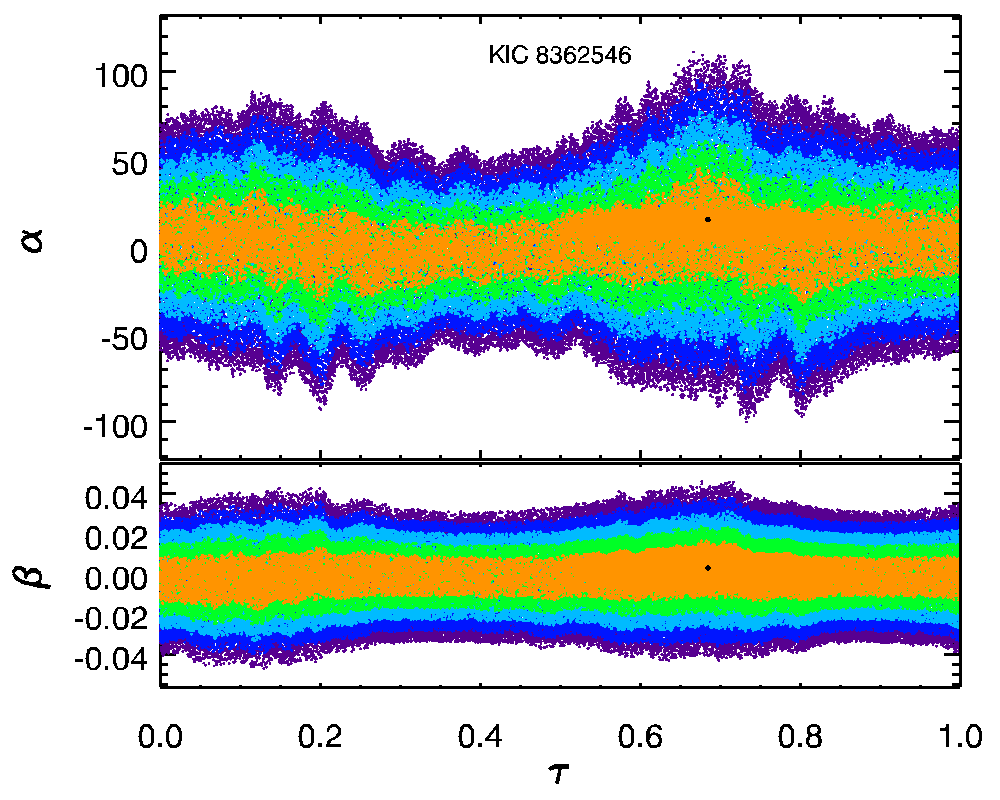}}
\end{tabular}
\caption{The $\chi^2$ map of all combinations of $\alpha, \tau, \beta$, projected on two planes, of the CP2 stars KIC~3326428 (left panel) and KIC~8386865 (middle panel), and the mCP star candidate KIC~8362546 (right panel). The black dot marks the location of the best solution. The colour scale (orange to violet) indicates the $\chi^2$ in standard deviation intervals (1 to 5 $\sigma$), while the white colour covers the space where the $\chi^2$ is more than 5$\sigma$ larger than the minimum or where there are no data (i.e., no parameter combinations have been computed). As expected, the results for the CP2 stars indicate antiphase variability in the UV and {\em Kepler} light curves. This is not the case for the mCP star candidate, which provides evidence that this star is not a mCP star after all.}
\label{fig:chi2}
\end{figure*}

\subsection{The asexual genetic algorithm}
\label{sec:aga}

For implementing the AGA, we described the NUV phase curve of a mCP star with the model:
\begin{equation}
\Delta m_{\rm NUV_{model}}(\phi) = \alpha \Delta m_{\rm Kep}(\phi + \tau) + \beta \, ,
\label{eq:nuvagamodel}
\end{equation}
where $\Delta m_{\rm Kep}$ is the {\em Kepler} average phase curve, $\Delta m_{\rm NUV_{model}}$ is the {\em GALEX} phase curve, while the three free parameters are: $\alpha$, the amplitude ratio of the NUV curve to the visible one; $\tau$, the phase difference (in the interval 0--1); and $\beta$, the offset between the mean of two curves.

The AGA will find the best $\alpha, \tau, \beta$ values by minimizing the reduced chi-square, defined as:
\begin{equation}
\chi_{\nu}^2 = \frac{1}{n-3}\sum_{i=1}^{n} \frac{ ( \Delta m_{{\rm NUV_{model}},i} - \Delta m_{{\rm NUV_{obs}},i} )^2}{\sigma^2_{{\rm NUV_{obs}},i} }
\label{eq:nuvagachi}
\end{equation}
where $\Delta m_{{\rm NUV_{obs}},i}$ is the value of the $i$-th point of the {\em GALEX} observed data, with error  $\sigma^2_{{\rm NUV_{obs}},i}$, $\Delta m_{{\rm NUV_{model}},i}$ is the value of the $i$-th point of the computed model of the {\em GALEX} phase curve, at the same $\phi$ as the observed point, and $n$ is the total number of points.

As a first step, for each free parameter ($\alpha, \tau, \beta$), an initial random population of $N_0$=3050 values (i.e. individuals) is generated in a defined interval and the model for the NUV phase curve (Equation~\ref{eq:nuvagamodel}) is computed. Then, the fitness of each individual is evaluated by means of a merit function, which in this case is the $\chi_{\nu}^2$ (Equation~\ref{eq:nuvagachi}). The $N_1$=50 individuals with the highest fitness (i.e., smaller $\chi_{\nu}^2$) are selected and passed to the next generation (this is the reason of the word "asexual" in the algorithm name). Then, for each free parameter, a new and smaller interval is defined, centered at the parameter value of each $N_1$ selected individual, where a new generation of $N_2$=60 random individuals is created. Again, the fitness of all $N_1 \times N_2 + N_1 (=N_0)$ individuals in this new generation is calculated and the process is iterated until a stopping criterion is achieved. Here, this criterion is that, for five successive generations, the $N_1$ individuals do not change and the difference between the minimum and maximum of their $\chi_{\nu}^2$ is less than $10^{-3}$.
Finally, the model (i.e., $\alpha, \tau$ and $\beta$) that has the minimum $\chi_{\nu}^2$ is assumed as the best solution and adopted as the NUV model of the phase curve of the star. \citet{canto2009} thoroughly described the algorithm and its performance.

We evaluated the error associated to the free parameters $\alpha, \tau$ and $\beta$ following the recipe of \citet{avni1976}. Using the $\chi^2$ values of individuals from all AGA iterations, we identified the volume, in the 3D parameter space, where the $\chi^2$ is lower than the ${\rm MIN}(\chi^2) + 3.52674$ (for 1$\sigma$ confidence level and 3 free parameters). The extreme values of the volume in each dimension provide the error on each parameter.
In order to further improve the error estimation, we added the $\chi^2$ from an extra set of random $\alpha, \tau$, $\beta$ combinations, more homogeneously distributed in the parameters space, which increased the total number of individuals to more than 1 million, on average. This procedure produces asymmetric statistical errors around the mean.

Figure~\ref{fig:chi2} illustrates, as an example, the $\chi^2$ maps of the three stars shown in Fig.~\ref{fig:phase}.
For the CP2 star KIC~3326428 (left panel), the parameters are well determined and constrained. The phase difference $\tau$ is almost 1 and the amplitude ratio is negative: this implies that the NUV and the visible curves are almost perfectly anticorrelated, which basically confirms the star as a CP2 star. (In fact, during the course of our study, we were able to confirm this candidate mCP star from \citet{hummerich2018} as a CP2 star by our own spectroscopic observations, cf. Section \ref{sec:cpsample}.)
For the CP2 star KIC~8386865, the presence of two maxima of similar amplitude during a rotational cycle is reflected in the presence of various relative minima also in the map, almost equidistant along $\tau$ and with a similar value of $| \alpha |$. Again, this implies that the NUV and the visible curves are anticorrelated, in expectations for this spectroscopically confirmed CP2 star. The latter is the most common result provided by AGA; in these cases, we only took into account the relative minimum where the best solution is located to estimate the errors on the parameters. Finally, in the case of the CP2 star candidate KIC~8362546, the AGA was not able to find a solution for the phase shift $\tau$ nor to significantly constrain the amplitude ratio $\alpha$, which can be viewed as evidence that the star is not a CP2 star (cf. the discussion in Section \ref{sec:notes_CP2candidates}).

\subsection{The fit with harmonic polynomials}
\label{sec:harmpol}

\citet{mikulasek2007} stated that the light curves of mCP stars can be well described by real-valued harmonic polynomials of 2nd degree. We adapted their definition for the case of the {\em Kepler} phase curves:
\begin{multline}
\Delta m_{\rm Kep_{model}}(\phi) = \overline{\Delta m_{\rm Kep}} + c_1 \cos(2\pi\phi) +  c_2 \sin(2\pi\phi) \\ + c_3 \cos(4\pi\phi) +c_4 \sin(4\pi\phi) \, .
\label{eq:hf}
\end{multline}
The free parameters in this equation are five: the four coefficients $c_i$ and the average $\overline{\Delta m_{\rm Kep}}$. We fitted the {\em Kepler} observations of all stars of the sample, using a robust least-squares minimization that makes use of the Levenberg-Marquardt algorithm.
We found that the best fits reproduce very well the shapes of the {\em Kepler} phase curves, even though the $\chi_{\nu}^2$ is, in many cases, quite large, because the {\em Kepler} observational errors are extremely small (the SNR ranges in the $\sim$2000-20,000 interval). 

Then, in order to reproduce a similar procedure that we used with the AGA fitting, we assumed the NUV phase curve to have the same shape as the {\em Kepler} curve. So, we defined the function:
\begin{equation}
\Delta m_{\rm NUV_{model}}(\phi) = \alpha_{\rm hf} \, \Delta m_{\rm Kep_{model}}(\phi+\tau_{\rm hf}) + \beta_{\rm hf}
\label{eq:fnuv}
\end{equation}
where the three free parameters $\alpha_{\rm hf}, \tau_{\rm hf}, \beta_{\rm hf}$ have the same meaning as in the AGA fit.
We then fitted the NUV observations, using the same least-squares minimization algorithm as before, to find the best fitting model of equation~\ref{eq:fnuv}.
As the number of free parameters is the same as in the AGA fitting, the $\chi_{\nu}^2$ values of the two methods are directly comparable.

\section{The results of the fitting procedure}
\label{sec:results}

\subsection{Results from the AGA}

\begin{figure*}
\centering
\begin{tabular}{ccc}
\resizebox{56mm}{!}{\includegraphics{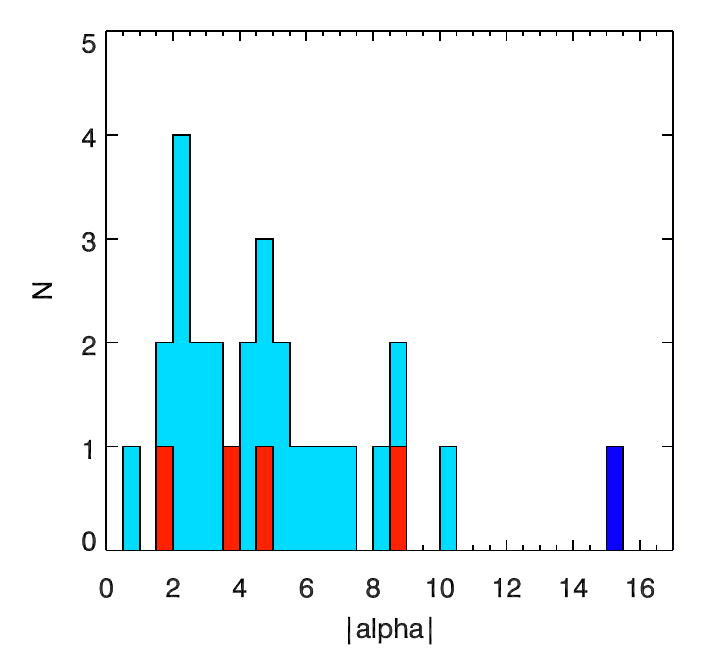}} &
\resizebox{56mm}{!}{\includegraphics{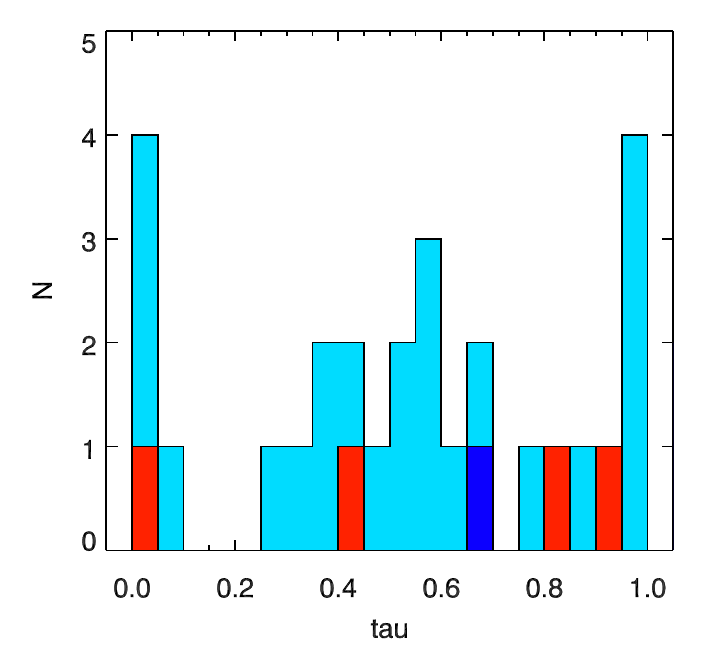}} &
\resizebox{56mm}{!}{\includegraphics{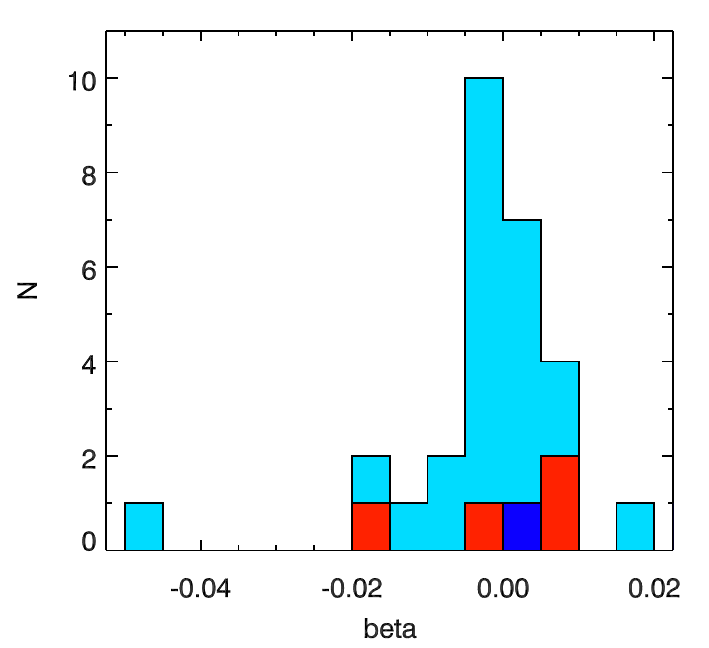}} 
\end{tabular}
\caption{Distributions of the best parameters provided by the AGA for the 28 sample stars. {\it Left panel:} distribution of the absolute value of $\alpha$ (amplitude ratio). {\it Central panel:} distribution of $\tau$ (phase difference). {\it Right panel:}  distribution of $\beta$ (magnitude offset). The light blue colour indicates the confirmed CP2 stars, the dark blue shows the candidate CP2 object, while the red displays non-CP2 stars.}
\label{fig:agaparameters}
\end{figure*}

We report the results provided by the AGA in Table~\ref{tab:agaresults}: the columns report the KIC ID number, the parameters ($\alpha$, $\beta$, $\tau$) of the best fit, along with their respective positive and negative 1$\sigma$ errors, 
the $\chi^2_\nu$ of the best fit, the number of {\em GALEX} points, the amplitude ($A$) of the {\em Kepler} light curve and the amplitude of the {\em GALEX} one, with its positive and negative 1$\sigma$ errors (see Sect.~\ref{sec:correlations} for the definition of the amplitude), the classification of the {\em Kepler} light curve and, finally, the Pearson's linear correlation coefficient. These aspects are discussed below.
We summarise these results in Fig.~\ref{fig:agaparameters}, where we show the distributions of the best solutions for the three free parameters. 

The distribution of the absolute values of $\alpha$ implies that the variability in the NUV is typically quite larger than in the visible: the average is $|\alpha|=5.0$ and only one star, KIC~9665384, has a slightly lower amplitude in the NUV ($\alpha=-0.95^{+2.75}_{-1.16}$), however it is one of the stars with the poorest fit (see below), as the error of the {\em GALEX} observations is quite high with respect to the flux variation (cf. also the corresponding phase plot in Fig. \ref{fig:appendixphasediagrams}).

The $\tau$ distribution is characterised by a peak around a phase difference of 0 (at both $\tau=0$ or 1). This means that the NUV curve is almost perfectly in phase (if $\alpha>0$) or antiphase (if $\alpha<0$) with the visible one. Another broader peak is centred at about $\tau=0.5$, which may also imply a strong positive or negative correlation, depending on the shape of the light curve.

As expected, the distribution of the brightness offset $\beta$ is centered around zero, but in many cases the value of $\beta$ is significant with respect to the NUV amplitude. This shift is due to the much lower number of data points in the {\em GALEX} light curves as compared to the {\em Kepler} ones.

\begin{figure}
\centering
\resizebox{\columnwidth}{!}{\includegraphics{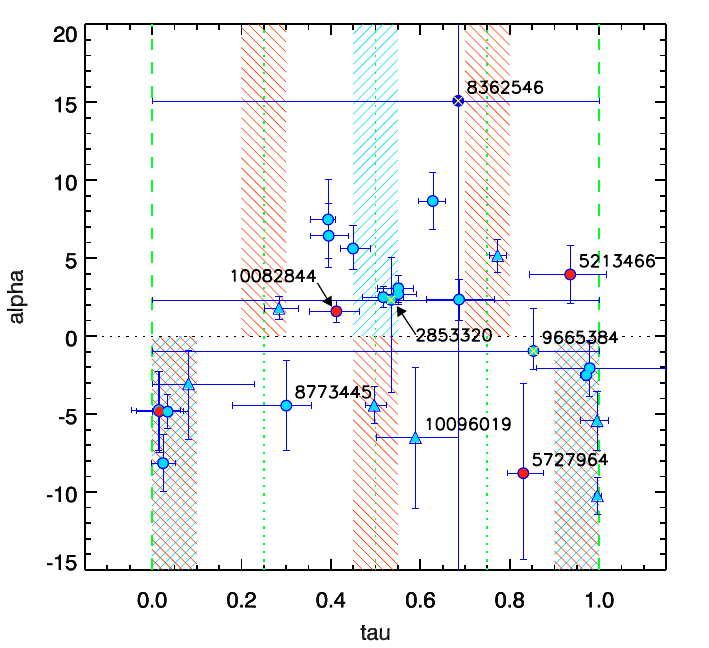}} 
\caption{Amplitude ratio $\alpha$ vs. phase difference $\tau$, as obtained with the AGA. The colour code is the same as in Fig.~\ref{fig:agaparameters}: the light blue symbols indicate the confirmed CP2 stars, the dark blue symbol marks the CP2 candidate (KIC~8362546), while the red symbols locate the non-CP2 stars. 
Stars with single-wave light curves are indicated by circles, while the triangles  mark the double-wave stars (see the classification in Table~\ref{tab:agaresults}). The three stars that have an undetermined value of $\tau$ are identified with a yellow cross on the symbol. The striped regions show the locations in the plane where a strong anticorrelation between the {\em Kepler} and the {\em GALEX} light curves is to be expected: cyan for single-wave light curves, red for the double-wave light curves. Some notable objects discussed in the text are identified with their KIC number.}
\label{fig:alphatau}
\end{figure}

In Fig.~\ref{fig:alphatau}, we present the plot of $\alpha$ vs. $\tau$. If the periodic variability in the visible range is anticorrelated with that in NUV, as we should expect for CP2 stars when the null wavelength is located somewhere in between the two intervals, we would expect the points to be located in specific places in the plot. These locations also depend on the distribution of the spots on the stellar surface that can produce single-wave or double-wave light curves. Assuming that the curves, at different wavelengths, have the same shape, a perfect anticorrelation is always obtained for $\tau=0$ or 1 and $\alpha < 0$. Then, in the case of single-wave light curves, a strong anticorrelation is also present when $\tau \sim 0.5$ and $\alpha > 0$. Regarding the stars with double-wave light curves, anticorrelation can also be achieved at $\tau \sim 0.5$, but for negative values of $\alpha$; furthermore, also the combinations of $\tau \sim 0.25$ and $\tau \sim 0.75$, with positive $\alpha$, may indicate anticorrelation. These regions are highlighted in Fig.~\ref{fig:alphatau}.
We have divided the sample in stars with single-wave or double-wave light curves (see Table~\ref{tab:agaresults}) and we observe in the plot that most of the stars lie in or not far from the correspondent regions of anticorrelation.

\begin{figure}
\centering
\resizebox{\columnwidth}{!}{\includegraphics{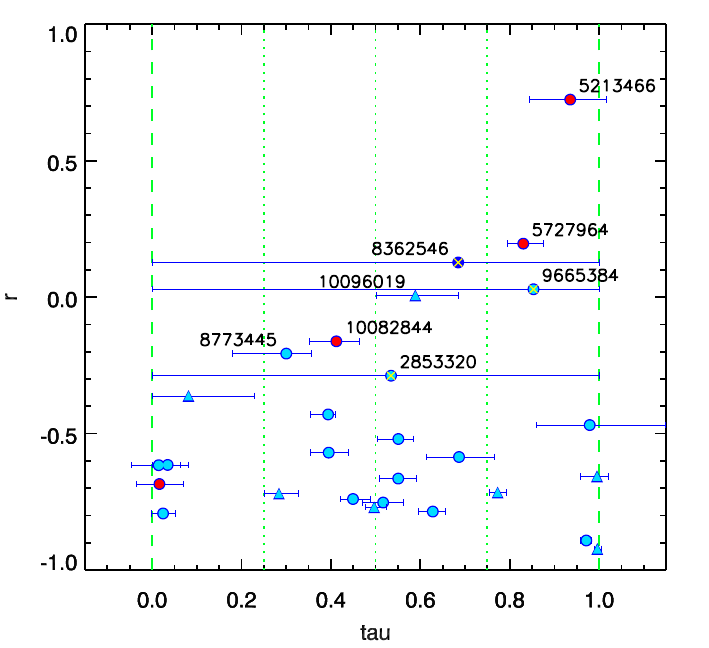}} 
\caption{The plot of the Pearson's linear coefficient $r$ vs. $\tau$. colours and symbols are the same as in Fig.~\ref{fig:alphatau}.}
\label{fig:pearsontau}
\end{figure}

There are, however, some notable exceptions that we discuss also with the help of Fig.~\ref{fig:pearsontau}, where we show the plot of the Pearson's linear coefficient $r$ that quantifies the correlation between the {\em GALEX} observed points and the smoothed {\em Kepler} light curve (at the values of $\tau$ of the NUV points).

First, there are three stars (the confirmed CP2 stars KIC~2853320 and 9665384 and the CP2 star candidate KIC~8362546) whose 1$\sigma$ error covers the entire $\tau$ valid range. We can therefore consider that the AGA left this parameter undetermined. These stars also have very large errors on $\alpha$ because the {\em GALEX} observational errors are very large with respect to the maximum flux variation. 
Furthermore, they show a poor correlation between the NUV and visible curves. 
Other four stars have $r \sim 0$: the non-CP stars KIC~10082844 and KIC~5727964 and the confirmed CP2 stars KIC~8773445 and KIC~1009601. For the three latter cases, the lack of correlation can be explained by the large NUV observational errors with respect to the amplitude of the flux variation and, in the case of KIC~5727964, also for the small number of {\em GALEX} data (11).
Regarding KIC~10082844, the small |$r$| is mainly caused by the large dispersion of the NUV flux values in the interval $0 < \phi <0.3$; this spread also produces a large $\chi^2_{\nu}$ of the AGA fit. Non-CP stars will be discussed in more detail in Section \ref{sec:notes_non-mCP}.

We also recall the fact that while for the majority of mCP stars, the null-wavelength region lies somewhere between the {\em GALEX} UV and {\em Kepler} passbands (and hence these stars are expected to show the corresponding anticorrelated light curve behaviour), for some mCP stars, the null wavelength is situated in the optical region, for example between the ZTF $g$ and $r$ filters \citep[e.g.][]{faltova21}.

Finally, the non-CP star KIC~5213466 shows a strong positive correlation between the {\em Kepler} and the {\em GALEX} NUV bands, with the only caveat that it is one of the stars with the smallest number of {\em GALEX} observations. This finding corroborates the result of \citet{hummerich2018} that this star is not an ACV variable. 

All the other stars show a clear anticorrelation between the flux of the {\em GALEX} NUV and {\em Kepler} wavelength intervals.

\begin{landscape}
\begin{table}
\caption{Results from the AGA.}
\label{tab:agaresults}
\begin{tabular}{lrrrrrrrrrrrrrrrrr}
\hline
KIC & $\alpha$ & $+\sigma_\alpha$ &  $-\sigma_\alpha$  &  $\tau$  & $+\sigma_\tau$  &  $-\sigma_\tau$  &  $\beta$ & $+\sigma_\beta$  &  $-\sigma_\beta$  &  $\chi^2_\nu$  &  n  &  $A_{\rm Kepler}$   &  $A_{\rm NUV}$  & $+\sigma_{A_{\rm NUV}}$   &  $-\sigma_{A_{\rm NUV}}$ &  wave  &  $r$ \\
 & & & & & & & (mmag) & (mmag) & (mmag) & & & (mmag) & (mmag) & (mmag) &  (mmag) & & \\
\hline
 2853320 &   2.33 &   2.74 &   5.90 & 0.535 & 0.465 & 0.535 &   -2.7 &    8.0 &    8.1 & 0.74 & 18 &  5.10 &  11.86 &  13.97 &  11.86 & 1 &  -0.29 \\
 2969628 &   3.06 &   0.86 &   0.91 & 0.551 & 0.034 & 0.047 &    1.4 &    3.5 &    3.7 & 2.63 & 21 &  5.99 &  18.37 &   5.13 &   5.43 & 1 &  -0.52 \\
 3326428 & -10.24 &   1.21 &   1.15 & 0.996 & 0.009 & 0.008 &  -49.8 &   12.2 &   11.2 & 3.48 & 18 & 16.65 & 170.58 &  19.22 &  20.15 & 2 &  -0.92 \\
 3945892 &   2.50 &   0.69 &   0.66 & 0.517 & 0.045 & 0.046 &    8.2 &    6.9 &    6.9 & 2.13 & 20 & 13.61 &  33.98 &   9.46 &   8.99 & 1 &  -0.75 \\
 4171302 &  -8.15 &   1.83 &   1.83 & 0.025 & 0.027 & 0.026 &   -5.7 &    5.2 &    5.4 & 2.28 & 18 &  4.34 &  35.37 &   7.95 &   7.93 & 1 &  -0.79 \\
 5000179 &  -4.43 &   1.20 &   1.19 & 0.497 & 0.026 & 0.019 &   -5.1 &   11.2 &   11.2 & 1.43 & 15 & 14.07 &  62.39 &  16.74 &  16.84 & 2 &  -0.77 \\
 5213466 &   3.95 &   1.85 &   1.87 & 0.935 & 0.081 & 0.093 &    5.9 &   20.0 &   17.9 & 1.17 & 11 & 12.81 &  50.63 &  23.75 &  23.95 & 1 &   0.72 \\
 5727964 &  -8.79 &   5.79 &   5.53 & 0.830 & 0.046 & 0.036 &    5.9 &   17.3 &   16.3 & 0.65 & 11 &  4.79 &  42.12 &  26.47 &  27.74 & 2 &   0.20 \\
 5739204 &  -2.06 &   1.81 &   1.83 & 0.979 & 0.168 & 0.121 &    0.0 &   18.9 &   18.5 & 1.04 & 18 & 16.76 &  34.45 &  30.68 &  30.38 & 1 &  -0.47 \\
 5774743 &   2.35 &   1.29 &   1.30 & 0.687 & 0.079 & 0.074 &    7.4 &    7.6 &    7.7 & 0.47 & 15 &  6.97 &  16.35 &   8.98 &   9.06 & 1 &  -0.58 \\
 6715809 &   8.66 &   1.86 &   1.83 & 0.628 & 0.028 & 0.032 &  -17.4 &   10.4 &   10.7 & 1.21 & 18 &  9.95 &  86.23 &  18.56 &  18.17 & 1 &  -0.79 \\
 6950556 &   1.78 &   0.75 &   0.73 & 0.284 & 0.042 & 0.032 &   -5.0 &    5.5 &    5.7 & 0.82 & 19 & 11.62 &  20.73 &   8.74 &   8.49 & 2 &  -0.72 \\
 7628336 &   7.48 &   2.57 &   2.47 & 0.394 & 0.016 & 0.041 &   -4.8 &    5.3 &    5.2 & 3.80 & 18 &  3.35 &  25.04 &   8.59 &   8.28 & 1 &  -0.43 \\
 7976845 &  -4.74 &   2.49 &   2.61 & 0.015 & 0.049 & 0.062 &   -0.1 &   36.1 &   35.9 & 1.29 & 18 & 24.83 & 117.70 &  64.86 &  61.74 & 1 &  -0.62 \\
 8362546 &  15.09 &  29.18 &  47.14 & 0.685 & 0.315 & 0.685 &    2.6 &   13.5 &   20.6 & 0.96 & 18 &  0.96 &  14.44 &  27.92 &  14.44 & 1 &   0.13 \\
 8386865 &   5.16 &   1.07 &   1.08 & 0.773 & 0.020 & 0.019 &   -2.8 &    5.0 &    4.9 & 2.95 & 16 &  6.35 &  32.77 &   6.80 &   6.84 & 2 &  -0.72 \\
 8569986 &  -4.82 &   2.58 &   2.63 & 0.016 & 0.054 & 0.052 &  -18.6 &   16.6 &   17.3 & 1.17 & 12 &  9.74 &  46.90 &  25.60 &  25.09 & 1 &  -0.68 \\
 8773445 &  -4.46 &   2.92 &   2.87 & 0.300 & 0.056 & 0.122 &  -14.3 &   16.0 &   15.9 & 0.61 & 17 &  9.68 &  43.19 &  27.77 &  28.28 & 1 &  -0.21 \\
 9541567 &  -2.50 &   0.29 &   0.32 & 0.971 & 0.011 & 0.013 &   -0.6 &    4.0 &    3.8 & 4.44 & 16 & 24.15 &  60.44 &   7.66 &   6.94 & 1 &  -0.89 \\
 9665384 &  -0.95 &   2.75 &   1.16 & 0.853 & 0.147 & 0.853 &   -3.5 &   11.8 &    8.8 & 0.55 & 16 & 15.60 &  14.86 &  18.05 &  14.86 & 1 &   0.03 \\
10082844 &   1.59 &   0.73 &   0.73 & 0.412 & 0.052 & 0.060 &   -4.1 &   10.7 &   10.7 & 4.23 & 17 & 18.00 &  28.59 &  13.11 &  13.06 & 1 &  -0.16 \\
10090722 &   6.44 &   2.06 &   2.05 & 0.395 & 0.044 & 0.041 &    4.4 &    6.5 &    6.6 & 2.41 & 15 &  4.37 &  28.12 &   8.98 &   8.97 & 1 &  -0.57 \\
10096019 &  -6.51 &   4.49 &   4.51 & 0.589 & 0.096 & 0.088 &   -1.1 &   12.5 &   12.4 & 0.86 & 22 &  3.75 &  24.42 &  16.94 &  16.84 & 2 &   0.01 \\
10685175 &   5.62 &   1.47 &   1.36 & 0.449 & 0.040 & 0.029 &    1.8 &    5.8 &    5.2 & 2.44 & 20 &  5.82 &  32.70 &   8.58 &   7.90 & 1 &  -0.74 \\
10905824 &  -3.06 &   2.17 &   3.57 & 0.081 & 0.147 & 0.081 &   -0.9 &    9.5 &    9.9 & 2.16 & 12 &  7.35 &  22.50 &  26.22 &  15.93 & 2 &  -0.36 \\
10959320 &  -5.43 &   1.92 &   1.88 & 0.995 & 0.026 & 0.037 &   15.4 &   12.0 &   12.2 & 2.50 & 17 & 10.57 &  57.38 &  19.82 &  20.26 & 2 &  -0.66 \\
11154043 &  -4.85 &   1.11 &   1.10 & 0.035 & 0.045 & 0.034 &    0.5 &    8.0 &    6.7 & 4.69 & 20 &  6.34 &  30.72 &   6.95 &   7.04 & 1 &  -0.62 \\
11465134 &   2.69 &   0.68 &   0.64 & 0.551 & 0.039 & 0.042 &    3.5 &    4.9 &    4.6 & 2.46 & 15 &  8.79 &  23.68 &   5.98 &   5.59 & 1 &  -0.66 \\
\hline
\end{tabular}
\end{table}
\end{landscape}

\subsection{Comparison with the results from the harmonic fitting}

\begin{figure}
\centering
\resizebox{\columnwidth}{!}{\includegraphics{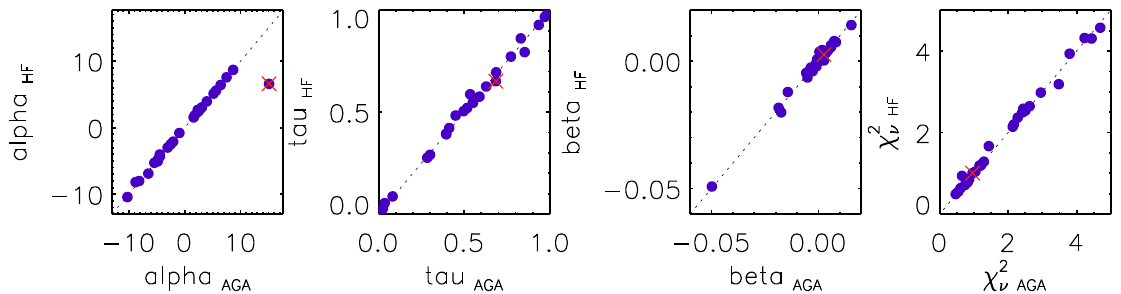}}
\caption{Comparison of the best fit parameters and the $\chi_{\nu}^2$ from AGA and harmonic fit. In all panels, the red cross indicates the location of the star KIC~8362546.}
\label{fig:comparison}
\end{figure}

The results of the harmonic fitting (HF) method are very similar to those obtained with the AGA, as shown in {\color{blue} Fig.~\ref{fig:comparison}}. The only significant difference is the value of $\alpha$ of the candidate CP2 star KIC~8362546, which changes from 15.1, in the AGA fit, to 6.6, with the HF; however, this star is one of the three objects that have a value of $\tau$ completely undetermined.

For some of the stars, some of the minimum $\chi_{\nu}^2$ obtained with the HF are slightly lower than the corresponding AGA values; this is possibly because the shape of the {\em Kepler} curve, used as a reference, is different in the two cases: for the AGA, it is a running average of the data points, while for the HF, it is the harmonic curve of Eq.~\ref{eq:hf}. 

Considering the very high level of agreement between the two fitting methods, in the rest of this work we only use the results from the AGA, as the assumption of equal shape for the visible and NUV light curves is more rigorous.

\subsection{Correlation of the light curve amplitudes with stellar parameters}
\label{sec:correlations}

Since the variability of CP2 stars is due to the different intensity of the line blanketing, at different wavelength intervals and with the phase of the rotation period, it is interesting to investigate if there is a correlation of the variability amplitude with the stellar parameters that affect the intensity of the absorption, in particular with effective temperature.

With that goal in mind, we adopt the main atmospheric parameters (the effective temperature $T_{\rm eff}$ and the surface gravity $\log{g}$) derived by using the calibrations of \citet{Paunzen2024}. These are based on four commonly used references 
\citep{2019AJ....158..138S,2022AandA...658A..91A,2023AandA...674A..28F,2023MNRAS.524.1855Z} of astrophysical parameters, which were then refined for the different subgroups of CP stars.
We report the $T_{\rm eff}$ and $\log{g}$ for the whole sample, along with their associated uncertainties, in Table~\ref{tab:deltaaparameters}: the sample of 28 stars covers, quite evenly, the $T_{\rm eff}$ range 7500--13000~K, where main-sequence A-type to late B-type objects are located, and the $\log{g}$ values are consistent with main-sequence objects. Note that the three stars with undetermined $\tau$ (the confirmed CP2 stars KIC~2853320 and KIC~9665384 and the CP2 star candidate KIC~8362546) are the hottest of the sample and the catalogue of \citet{mathur2017} places them as the farthest of the sample.

\begin{figure}
\centering
\resizebox{0.9\columnwidth}{!}{\includegraphics{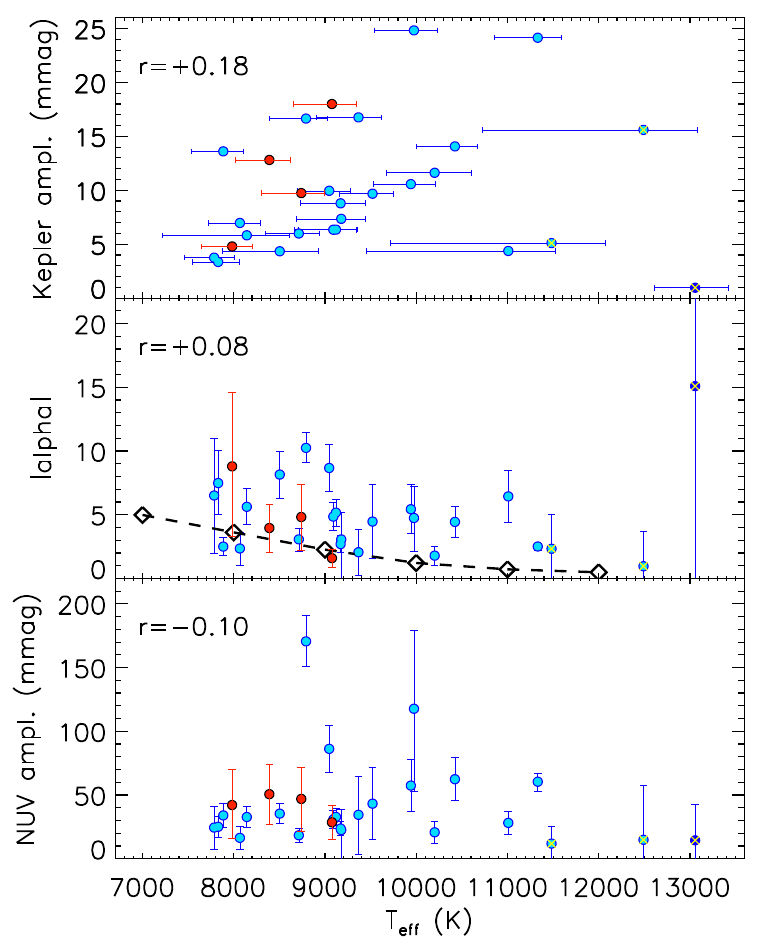}}
\caption{Plots of the {\em Kepler} light-curve amplitude (upper panel), the absolute value of the amplitude ratio $\alpha$ (middle panel) and the NUV light-curve amplitude (lower panel) vs. $T_{\rm eff}$. In the upper panel, the amplitude error is omitted, while in the middle and lower panels, for the sake of clarity, the error on $T_{\rm eff}$ is not plotted. Light blue symbols indicate confirmed CP2 stars, the dark blue one displays the CP2 candidate, while the red symbols show the non-CP2 stars. The three stars that have an undetermined value of $\tau$ are identified with a yellow cross on the symbol.
In the middle panel, the diamonds mark the locations of the theoretical $\alpha$ values obtained with synthetic spectra, as explained in Sect.~\ref{sec:modatm}. In each panel the linear correlation coefficient value is also indicated.}
\label{fig:amplitudeteff}
\end{figure}

We define the amplitude of the optical variability as half the difference between the maximum and the minimum values of the smoothed {\em Kepler} phase curve, while the NUV amplitude is obtained by multiplying the optical amplitude by $\alpha$.
In Fig.~\ref{fig:amplitudeteff}, we show the plot of the optical amplitude, the amplitude ratio $\alpha$, and the NUV amplitude with respect to the $T_{\rm eff}$. 
The Pearson's linear correlation coefficient $r$, calculated excluding the 4 non-CP2 stars and indicated in each panel, shows that there is no correlation between each of the amplitudes and $T_{\rm eff}$. This result does not change significantly if we also exclude the three stars with undetermined $\tau$: only the correlation between $T_{\rm eff}$ and the amplitude of the {\em Kepler} variability becomes mildly positive ($r=+0.51$), while the $T_{\rm eff}$ correlation with the other two quantities remains insignificant ($|r|<0.23$).
The lack of significant correlation of the amplitudes also stands with $\log{g}$: in all cases $|r|<0.22$. Likewise, no correlation is found with the radius and the mass ($|r|<0.23$),  taken from the \citet{mathur2017} catalogue.

We also searched for correlations with stellar age. \citet{berger2020} determined the age of a large number of stars in the {\em Kepler} prime field, collecting photometric and spectroscopic data from several sources and the parallaxes from Gaia DR2 \citep[e.g.,][]{gaiacollaboration2018} and making use of the isochrone fitting method. From their catalogue, we acquired the age and the fraction of the age on the main sequence for 23 stars.
We computed $r$ for both age-related quantities versus the {\em Kepler} amplitude, the NUV amplitude and their ratio $\alpha$: in no case the value of $|r|$ is larger than 0.33, indicating, once again, that there is no significant correlation between these quantities.

\begin{table}
\caption{Effective temperatures and surface gravities of the sample stars, derived using the calibration of \citet{Paunzen2024}.}
\label{tab:deltaaparameters}
\begin{tabular}{rrrrr}
\hline
KIC & $T_{\rm eff}$ & $\sigma_{T_{\rm eff}}$ & $\log{g}$ & $\sigma_{\log{g}}$ \\
    & (K) & (K) & (dex) & (dex)\\
\hline
2853320  & 11483 & 1427 &  4.075 &  0.181  \\
2969628  &  8713 &  119 &  3.989 &  0.023  \\
3326428  &  8792 &  833 &  3.872 &  0.048  \\
3945892  &  7885 &  306 &  4.014 &  0.070  \\
4171302  &  8504 &  469 &  4.052 &  0.088  \\
5000179  & 10424 &  902 &  3.992 &  0.071  \\
5213466  &  8390 &  240 &  3.913 &  0.110  \\
5727964  &  7982 &  176 &  3.995 &  0.047  \\
5739204  &  9367 &  325 &  4.047 &  0.037  \\
5774743  &  8067 &  150 &  3.946 &  0.048  \\
6715809  &  9046 &  320 &  4.207 &  0.036  \\
6950556  & 10202 &  390 &  3.990 &  0.077  \\
7628336  &  7829 &  233 &  4.029 &  0.045  \\
7976845  &  9975 & 1469 &  4.057 &  0.135  \\
8362546  & 13057 & 1404 &  4.138 &  0.081  \\
8386865  &  9120 &   85 &  4.038 &  0.044 \\
8569986  &  8740 &  194 &  4.048 &  0.053 \\
8773445  &  9521 &  495 &  3.993 &  0.034  \\
9541567  & 11332 & 1788 &  3.949 &  0.088  \\
9665384  & 12491 &  566 &  4.005 &  0.172  \\
10082844 &  9076 &  502 &  4.041 &  0.159  \\
10090722 & 11009 & 1258 &  3.785 &  0.096  \\
10096019 &  7785 &   32 &  4.067 &  0.078  \\
10685175 &  8143 &  192 &  4.179 &  0.103 \\
10905824 &  9178 &  238 &  4.136 &  0.085 \\
10959320 &  9941 &  777 &  4.037 &  0.116  \\
11154043 &  9091 &  326 &  3.949 &  0.046  \\
11465134 &  9171 &  383 &  3.906 &  0.078 \\
\hline
\end{tabular}
\end{table}

\subsubsection{Exploring the lack of correlation with theoretical models}
\label{sec:modatm}

Since, as we already said, the variability of the CP2 stars depends on the differential line blanketing on the stellar surface, the lack of significant correlations between the amplitude of the light curves and the investigated stellar parameters (especially $T_{\rm eff}$) is quite puzzling. 

\begin{figure*}
\centering
\resizebox{0.9\textwidth}{!}{\includegraphics{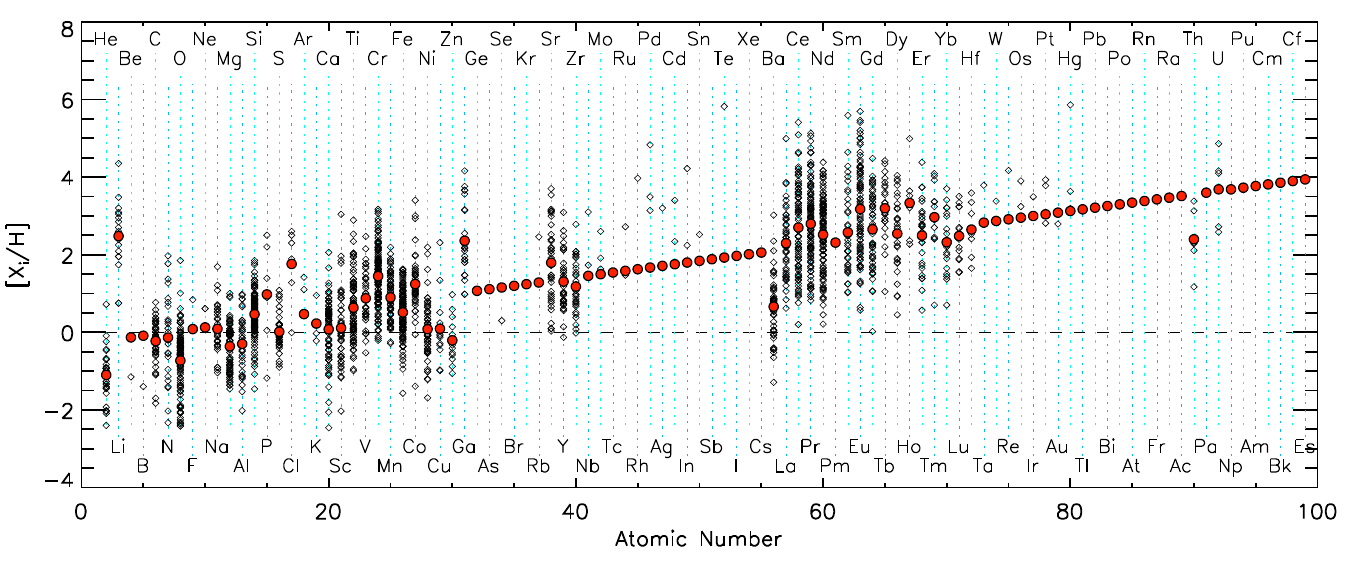}}
\caption{Abundances of the CP2 stars in the catalogue of \citet{ghazaryan2018} (black diamonds). The red dots indicate the adopted abundance values for a generic CP2 star, as explained in the text.} 
\label{fig:ghar}
\end{figure*}

Aimed at understanding the reasons behind that absence of correlations, we conducted an experiment using a small library of model atmospheres and theoretical spectral energy distributions in the $T_{\rm eff}$ range 7000--12000~K, compatible with our target parameters. We considered the case in which the variability is due to the presence of a spot, with typical abundances of a CP2 star, on the surface of a star with otherwise solar chemical composition.\footnote{We note that the bulk chemical composition
of CP2 stars is unknown and therefore the choice of an appropriate chemical composition for theoretical tracks remains an
open question (cf. the discussion in \citealt{bagnulo06}). However, if atomic diffusion is accepted as the main mechanism for producing the observed chemical peculiarities, the assumption of an overall abundance close to solar seems reasonable (cf. also \citealt{huemmerich20}).}

We first construct a chemical composition that may represent a `generic CP2 star'. For this type of stars, we extracted the spectroscopic abundance of individual chemical elements, relative to the Sun, from the catalogue of \citet{ghazaryan2018}. The total number of CP2 stars in this catalogue is 188. 71 chemical elements have measurements, but 25 elements have no more than three values, while six elements have more than 100 data points (O, Si, Cr, Fe, Pr, Nd). For each chemical element, we computed the mean and the standard deviation of the abundance values. 

For the average chemical composition, we adopted the mean value of each element with four or more measurements or the value of a linear fit for the other elements, as shown in Fig.~\ref{fig:ghar}.
These abundances are lower than in the Sun for just 9 elements, most notably for He and O, but also for Be, B, C, N, Mg, Al and Zn.
As the reference solar chemical composition, we considered the abundances of \citet{asplund2021}.

\begin{figure}
\centering
\resizebox{\columnwidth}{!}{\includegraphics{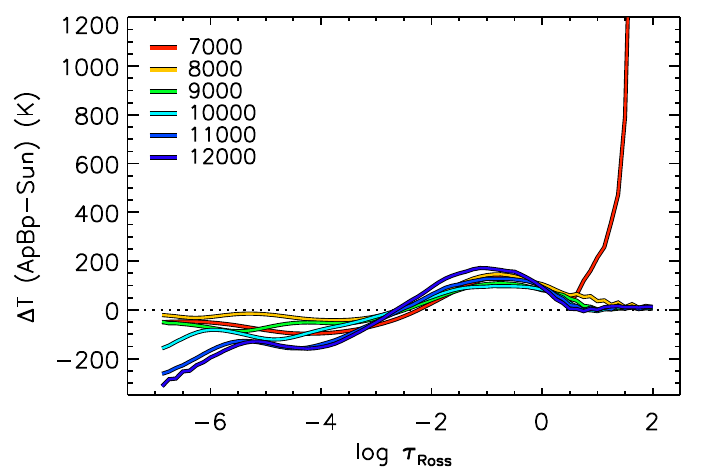}}
\resizebox{\columnwidth}{!}{\includegraphics{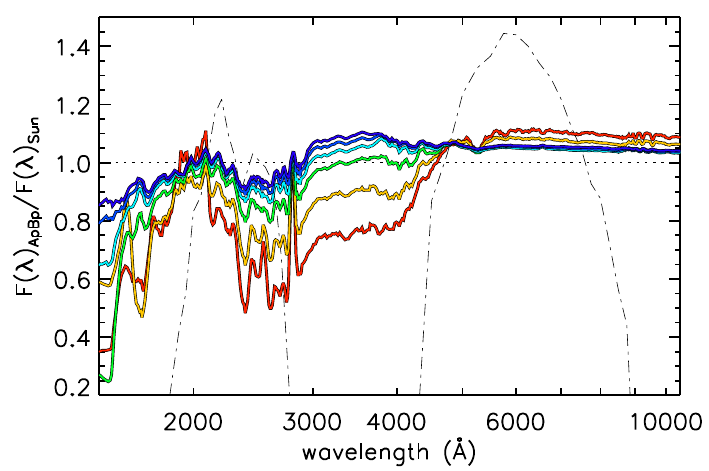}}
\caption{{\bf Upper panel}: profiles of the temperature difference between model atmospheres with CP2 and solar chemical composition, at each $T_{\rm eff}$. The colour scale is indicated in the plot. The 7000~K model has the largest difference of 3485~K at $\log{\tau_{\rm Ross}}=1.875$. {\bf Lower panel}: surface flux ratio between the CP2 and the solar theoretical SEDs, which were smoothed with a five-point boxcar. The colour scale is the same as in the upper panel. The dash-dotted lines represent the {\em GALEX} NUV and the {\em Kepler} filter responses, in an arbitrary scale.}
\label{fig:seds}
\end{figure}

We then computed, using the Fortran code DFSYNTHE\footnote{\url{http://wwwuser.oats.inaf.it/castelli/sources/dfsynthe.html}} by Robert Kurucz \citep[see, e.g.,][]{castelli2005}, the opacity distribution functions (ODFs; i.e. the pre-computed library of opacity for suitable intervals in temperature, pressure, and microturbulent velocity) for the solar and for the generic CP2 chemical compositions.
In this way, we were able to quickly compute the model atmospheres and SEDs for both chemical compositions, a fixed $\log{g}=4$~dex and $T_{\rm eff} = 7000, 8000, 9000, 10000, 11000, 12000$~K with Kurucz's ATLAS9 code \citep[e.g.,][]{kurucz2005}. Convection is described by the mixing-length theory \citep[MLT;][]{bohmvitense1958}; to treat its limited contribution to energy transport in A-type stars, we follow the indication of \citet{smalley2004} to adopt the MLT parameter $\alpha = l/H_p = 0.5$. The models have 72 layers and the SEDs are computed in 1221 wavelength intervals.
In Fig.~\ref{fig:seds}, we present the differences of the temperature profiles between the CP2 and the solar model atmospheres and their ratio of the surface flux, for each $T_{\rm eff}$.
The hotter layers, at $\log{\tau_{\rm Ross}} > 0.5$, of the 7000 K CP2 model are caused by a lower level of convective flux, producing a steeper temperature gradient, while the models at warmer $T_{\rm eff}$ have quite similar temperatures in the optical thick regime. All the CP2 models are slightly warmer than the corresponding solar ones in the range $-2 \lesssim \log{\tau_{\rm Ross}} \lesssim 0$, while they become colder in the upper layers.
 
The low-resolution theoretical SEDs (Fig.~\ref{fig:seds}) show that, with respect to the solar reference, the CP2 models produce lower emission in the {\em GALEX} NUV passband over the whole $T_{\rm eff}$ interval. This depression is compensated by a positive excess in the optical regime and all through the {\em Kepler} passband.
This results imply that in our scenario (a spot with CP2 abundances on a star with solar chemical composition), the theoretical {\em GALEX} NUV and {\em Kepler} light curves are anticorrelated. Furthermore, since in the NUV the flux decreases more strongly, with decreasing $T_{\rm eff}$, than it increases in the visible, we expect a significant dependence on $T_{\rm eff}$.

In order to quantify this dependence, we constructed synthetic light curves at each $T_{\rm eff}$. We described the time-invariant distribution of spots with CP2 abundances by a filling factor $\omega(\phi)$, that indicates the fraction of the observed surface occupied by the spots and depends on the phase $\phi$. Therefore, the flux can be written as $ f = \omega f_{\rm CP2} + (1 - \omega) f_\odot$. While the amplitude of the light curve in a single passband depends on $\omega(\phi)$, the amplitude ratio $\alpha$ does not. 

We measured $\alpha$ between the synthetic NUV and the {\em Kepler} light curves at all $T_{\rm eff}$. The results confirm that $\alpha$ strongly depends on $T_{\rm eff}$: as shown in the middle panel of Fig.~\ref{fig:amplitudeteff} by the black diamonds, the value of $|\alpha|$ smoothly decreases with $T_{\rm eff}$; it has a maximum value of $|\alpha| = 4.98$ at 7000~K, while it becomes lower than 1 for the two hotter models (0.71 at 11000~K and 0.48 at 12000~K).

We note that the vast majority of the stars in our sample have $|\alpha|$ larger than the theoretical value from our experiment. This suggests that the assumed chemical compositions in our scenario are inadequate to describe the average properties of the sample.

The experiment also shows that, at fixed abundances, there is a clear correlation between the light curve amplitude ratios and $T_{\rm eff}$, while our sample shows neither a correlation with this quantity nor with the other stellar properties that we explored.
The data from the catalogue of \citet{ghazaryan2018} show a large dispersion among the CP2 stars, sometimes up to four orders of magnitude, in the abundance of individual elements. Therefore, each individual chemical composition may be quite different from one star to the other, and it is the dominant factor that generates the $\alpha$ value, as it determines the line blanketing in the different wavelength bands.

\section{Notes on single objects}
\label{sec:notes}

This section briefly discusses some noteworthy objects.

\subsection{The CP2 star candidates KIC~2969628, 3326428 and 8362546}
\label{sec:notes_CP2candidates}

Initially, three candidate CP2 stars from \citet{hummerich2018} entered our sample (KIC~2969628, 3326428 and 8362546). Two of them, KIC~2969628 and 3326428 clearly show anticorrelation between the NUV and visible light curves, which points to these objects being indeed CP2 stars. At a subsequent point in our study, we were able to confirm both objects as classical CP2 stars by our own spectroscopic observations (cf. Section \ref{sec:cpsample}), which shows that the assumptions drawn from the variability pattern have been valid.

On the other hand, the mCP star candidate KIC~8362546 does not show any correlation between the NUV and visible light curves and has an undetermined value of $\tau$, which favours the scenario that this star is not a mCP star or that the null wavelength region is located at bluer wavelengths than the {\em GALEX} NUV passband. To distinguish between the two hypotheses, spectroscopic or far-UV observations are needed.

\subsection{The non-mCP stars KIC~5213466, 5727964, 8569986, and 10082844}
\label{sec:notes_non-mCP}

The four non-CP stars in our sample should not show the distinctive anti-correlation between the NUV and visible light curves as ACV variables do. In agreement with this expectation, the non-CP star KIC~5213466 (spectral type A1 V; \citealt{hummerich2018}) shows a strong positive correlation between both wavelength bands. It is, however, one of the stars with the minimum number of GALEX observations. Nevertheless, this finding corroborates the result of \citet{hummerich2018} that this star is not an ACV star. Its {\em Kepler} light curve shows large variations in amplitude and shape which are not expected in this type of variables. This is also true for the non-CP stars KIC~5727964 (A6 V; \citealt{hummerich2018}) and KIC~10082844 (A0 V; \citealt{hummerich2018}), which, in agreement with this, do not show any significant correlation between the NUV and visible light curves. However, the amplitudes of the variability are small in both objects and the NUV errors quite large.
Lastly, the non-CP star KIC~8569986 (A2 V; \citealt{hummerich2018}) has a significant Pearson's $r=-0.68$; 
nevertheless, the low number of the {\em GALEX} points, their dispersion and a large phase gap in its phase diagram (see Appendix~\ref{sec:appendixa}) cast doubts on the real presence of an anticorrelation. A campaign of simultaneous photometric visible and ultraviolet (preferably in the far-UV) observations and/or high-resolution spectra of this star would be needed to clarify the nature of this star.

\section{Conclusions}
\label{sec:conclusions}

Based on observational data from the GALEX and {\em Kepler} prime missions, we carried out a study of the properties of the photometric variability of a sample of 22 spectroscopically confirmed CP2 stars, one photometrically confirmed mCP star (KIC 7976845), one mCP star candidate (KIC 8362546), and four non-CP stars (KIC 5213466, 5727964, 8569986, 10082844) in the NUV and visible wavelength regions. To date, this is the largest sample of mCP stars studied in the NUV wavelength region.  We furthermore investigated the presence of a correlation of the variability amplitudes in both wavelength regions with stellar parameters such as effective temperature and surface gravity. To connect our findings to theoretical considerations, we calculated model atmospheres, spectral energy distribution profiles and synthetic light curves.

The main findings are summarised in the following:

\begin{itemize}
    \item We observe antiphase variations between the NUV and optical light curves in the majority of mCP stars. This indicates that the presence of this particular variability pattern is common also in the NUV wavelength interval and not only at FUV wavelengths. It also means that the combination of NUV and visible observations are suitable for identifying mCP star candidates.
     
    \item While the theoretical calculations show that, at fixed abundances, a clear correlation between the light curve amplitude ratios and $T_{\rm eff}$ is expected, our sample does not show a correlation with any of the investigated properties. This may be due to the highly individualistic abundance patterns of CP2 stars, which are the main contributors to the line blanketing in different wavelength bands. 
\end{itemize}

\section*{Acknowledgements}

This research has made use of the SIMBAD database, operated at CDS, Strasbourg, France.

\section*{Data Availability}

The {\em GALEX} NUV data, used in this paper, will be published in a future publication, but can be requested in advance to the first author. The {\em Kepler} light curves and the LAMOST spectra are publicly available online.



\bibliographystyle{mnras}
\bibliography{./manuscript} 

\begin{thebibliography}{}
\makeatletter
\relax
\def\mn@urlcharsother{\let\do\@makeother \do\$\do\&\do\#\do\^\do\_\do\%\do\~}
\def\mn@doi{\begingroup\mn@urlcharsother \@ifnextchar [ {\mn@doi@}
  {\mn@doi@[]}}
\def\mn@doi@[#1]#2{\def\@tempa{#1}\ifx\@tempa\@empty \href
  {http://dx.doi.org/#2} {doi:#2}\else \href {http://dx.doi.org/#2} {#1}\fi
  \endgroup}
\def\mn@eprint#1#2{\mn@eprint@#1:#2::\@nil}
\def\mn@eprint@arXiv#1{\href {http://arxiv.org/abs/#1} {{\tt arXiv:#1}}}
\def\mn@eprint@dblp#1{\href {http://dblp.uni-trier.de/rec/bibtex/#1.xml}
  {dblp:#1}}
\def\mn@eprint@#1:#2:#3:#4\@nil{\def\@tempa {#1}\def\@tempb {#2}\def\@tempc
  {#3}\ifx \@tempc \@empty \let \@tempc \@tempb \let \@tempb \@tempa \fi \ifx
  \@tempb \@empty \def\@tempb {arXiv}\fi \@ifundefined
  {mn@eprint@\@tempb}{\@tempb:\@tempc}{\expandafter \expandafter \csname
  mn@eprint@\@tempb\endcsname \expandafter{\@tempc}}}

\bibitem[\protect\citeauthoryear{{Anders} et~al.,}{{Anders}
  et~al.}{2022}]{2022AandA...658A..91A}
{Anders} F.,  et~al., 2022, \mn@doi [\aap] {10.1051/0004-6361/202142369}, \href
  {https://ui.adsabs.harvard.edu/abs/2022A&A...658A..91A} {658, A91}

\bibitem[\protect\citeauthoryear{{Asplund}, {Amarsi}  \& {Grevesse}}{{Asplund}
  et~al.}{2021}]{asplund2021}
{Asplund} M.,  {Amarsi} A.~M.,   {Grevesse} N.,  2021, \mn@doi [\aap]
  {10.1051/0004-6361/202140445}, \href
  {https://ui.adsabs.harvard.edu/abs/2021A&A...653A.141A} {653, A141}

\bibitem[\protect\citeauthoryear{{Auri{\`e}re} et~al.,}{{Auri{\`e}re}
  et~al.}{2007}]{auriere07}
{Auri{\`e}re} M.,  et~al., 2007, \mn@doi [\aap] {10.1051/0004-6361:20078189},
  \href {http://adsabs.harvard.edu/abs/2007A%26A...475.1053A} {475, 1053}

\bibitem[\protect\citeauthoryear{{Avni}}{{Avni}}{1976}]{avni1976}
{Avni} Y.,  1976, \mn@doi [\apj] {10.1086/154870}, \href
  {https://ui.adsabs.harvard.edu/abs/1976ApJ...210..642A} {210, 642}

\bibitem[\protect\citeauthoryear{{Babcock}}{{Babcock}}{1947}]{babcock47}
{Babcock} H.~W.,  1947, \mn@doi [\apj] {10.1086/144887}, \href
  {http://adsabs.harvard.edu/abs/1947ApJ...105..105B} {105, 105}

\bibitem[\protect\citeauthoryear{{Babusiaux} et~al.,}{{Babusiaux}
  et~al.}{2022}]{GAIA2}
{Babusiaux} C.,  et~al., 2022, arXiv e-prints, \href
  {https://ui.adsabs.harvard.edu/abs/2022arXiv220605989B} {p. arXiv:2206.05989}

\bibitem[\protect\citeauthoryear{{Bagnulo}, {Landstreet}, {Mason}, {Andretta},
  {Silaj}  \& {Wade}}{{Bagnulo} et~al.}{2006}]{bagnulo06}
{Bagnulo} S.,  {Landstreet} J.~D.,  {Mason} E.,  {Andretta} V.,  {Silaj} J.,
  {Wade} G.~A.,  2006, \mn@doi [\aap] {10.1051/0004-6361:20054223}, \href
  {https://ui.adsabs.harvard.edu/abs/2006A&A...450..777B} {450, 777}

\bibitem[\protect\citeauthoryear{{Balona}}{{Balona}}{2017}]{balona17}
{Balona} L.~A.,  2017, \mn@doi [\mnras] {10.1093/mnras/stx265}, \href
  {https://ui.adsabs.harvard.edu/abs/2017MNRAS.467.1830B} {467, 1830}

\bibitem[\protect\citeauthoryear{{Bauer-Fasching} et~al.,}{{Bauer-Fasching}
  et~al.}{2024}]{Bauer-Fasching2024}
{Bauer-Fasching} B.,  et~al., 2024, \mn@doi [\aap]
  {10.1051/0004-6361/202347476}, \href
  {https://ui.adsabs.harvard.edu/abs/2024A&A...687A.211B} {687, A211}

\bibitem[\protect\citeauthoryear{{Bellm} et~al.,}{{Bellm} et~al.}{2019}]{ZTF1}
{Bellm} E.~C.,  et~al., 2019, \mn@doi [\pasp] {10.1088/1538-3873/aaecbe}, \href
  {https://ui.adsabs.harvard.edu/abs/2019PASP..131a8002B} {131, 018002}

\bibitem[\protect\citeauthoryear{{Berger}, {Huber}, {van Saders}, {Gaidos},
  {Tayar}  \& {Kraus}}{{Berger} et~al.}{2020}]{berger2020}
{Berger} T.~A.,  {Huber} D.,  {van Saders} J.~L.,  {Gaidos} E.,  {Tayar} J.,
  {Kraus} A.~L.,  2020, \mn@doi [\aj] {10.3847/1538-3881/159/6/280}, \href
  {https://ui.adsabs.harvard.edu/abs/2020AJ....159..280B} {159, 280}

\bibitem[\protect\citeauthoryear{{Bertone}, {Sachkov}, {Olmedo}, {Olmedo}  \&
  {Chavez}}{{Bertone} et~al.}{2020}]{bertone2020}
{Bertone} E.,  {Sachkov} M.,  {Olmedo} D.,  {Olmedo} M.,   {Chavez} M.,  2020,
  in {Neiner} C.,  {Weiss} W.~W.,  {Baade} D.,  {Griffin} R.~E.,  {Lovekin}
  C.~C.,   {Moffat} A.~F.~J.,  eds, Stars and their Variability Observed from
  Space. pp 199--200

\bibitem[\protect\citeauthoryear{{Bianchi}, {Shiao}  \& {Thilker}}{{Bianchi}
  et~al.}{2017}]{bianchi2017}
{Bianchi} L.,  {Shiao} B.,   {Thilker} D.,  2017, \mn@doi [\apjs]
  {10.3847/1538-4365/aa7053}, \href
  {https://ui.adsabs.harvard.edu/abs/2017ApJS..230...24B} {230, 24}

\bibitem[\protect\citeauthoryear{{B{\"o}hm-Vitense}}{{B{\"o}hm-Vitense}}{1958}]{bohmvitense1958}
{B{\"o}hm-Vitense} E.,  1958, \zap, \href
  {https://ui.adsabs.harvard.edu/abs/1958ZA.....46..108B} {46, 108}

\bibitem[\protect\citeauthoryear{{Borucki} et~al.,}{{Borucki}
  et~al.}{2008}]{Borucki08}
{Borucki} W.,  et~al., 2008, in {Sun} Y.-S.,  {Ferraz-Mello} S.,   {Zhou}
  J.-L.,  eds,  IAU Symposium Vol. 249, Exoplanets: Detection, Formation and
  Dynamics. pp 17--24, \mn@doi{10.1017/S174392130801630X}

\bibitem[\protect\citeauthoryear{{Borucki} et~al.,}{{Borucki}
  et~al.}{2010}]{borucki2010}
{Borucki} W.~J.,  et~al., 2010, \mn@doi [Science] {10.1126/science.1185402},
  \href {https://ui.adsabs.harvard.edu/abs/2010Sci...327..977B} {327, 977}

\bibitem[\protect\citeauthoryear{{Brown}, {Latham}, {Everett}  \&
  {Esquerdo}}{{Brown} et~al.}{2011}]{brown2011}
{Brown} T.~M.,  {Latham} D.~W.,  {Everett} M.~E.,   {Esquerdo} G.~A.,  2011,
  \mn@doi [\aj] {10.1088/0004-6256/142/4/112}, \href
  {http://adsabs.harvard.edu/abs/2011AJ....142..112B} {142, 112}

\bibitem[\protect\citeauthoryear{{Cant{\'o}}, {Curiel}  \&
  {Mart{\'\i}nez-G{\'o}mez}}{{Cant{\'o}} et~al.}{2009}]{canto2009}
{Cant{\'o}} J.,  {Curiel} S.,   {Mart{\'\i}nez-G{\'o}mez} E.,  2009, \mn@doi
  [\aap] {10.1051/0004-6361/200911740}, \href
  {https://ui.adsabs.harvard.edu/abs/2009A&A...501.1259C} {501, 1259}

\bibitem[\protect\citeauthoryear{{Carrasco} et~al.,}{{Carrasco}
  et~al.}{2021}]{carrasco21}
{Carrasco} J.~M.,  et~al., 2021, \mn@doi [\aap] {10.1051/0004-6361/202141249},
  \href {https://ui.adsabs.harvard.edu/abs/2021A&A...652A..86C} {652, A86}

\bibitem[\protect\citeauthoryear{{Castelli}}{{Castelli}}{2005}]{castelli2005}
{Castelli} F.,  2005, Memorie della Societa Astronomica Italiana Supplementi,
  \href {https://ui.adsabs.harvard.edu/abs/2005MSAIS...8...34C} {8, 34}

\bibitem[\protect\citeauthoryear{{Conroy}, {Pr{\v{s}}a}, {Stassun}, {Orosz},
  {Fabrycky}  \& {Welsh}}{{Conroy} et~al.}{2014}]{conroy14}
{Conroy} K.~E.,  {Pr{\v{s}}a} A.,  {Stassun} K.~G.,  {Orosz} J.~A.,  {Fabrycky}
  D.~C.,   {Welsh} W.~F.,  2014, \mn@doi [\aj] {10.1088/0004-6256/147/2/45},
  \href {https://ui.adsabs.harvard.edu/abs/2014AJ....147...45C} {147, 45}

\bibitem[\protect\citeauthoryear{Cui et~al.,}{Cui et~al.}{2012}]{lamost2}
Cui X.-Q.,  et~al., 2012, Research in Astronomy and Astrophysics, 12, 1197

\bibitem[\protect\citeauthoryear{{Faltov{\'a}} et~al.,}{{Faltov{\'a}}
  et~al.}{2021}]{faltova21}
{Faltov{\'a}} N.,  et~al., 2021, \mn@doi [\aap] {10.1051/0004-6361/202141534},
  \href {https://ui.adsabs.harvard.edu/abs/2021A&A...656A.125F} {656, A125}

\bibitem[\protect\citeauthoryear{{Fouesneau} et~al.,}{{Fouesneau}
  et~al.}{2023}]{2023AandA...674A..28F}
{Fouesneau} M.,  et~al., 2023, \mn@doi [\aap] {10.1051/0004-6361/202243919},
  \href {https://ui.adsabs.harvard.edu/abs/2023A&A...674A..28F} {674, A28}

\bibitem[\protect\citeauthoryear{{Gaia Collaboration} et~al.,}{{Gaia
  Collaboration} et~al.}{2016}]{GAIA1}
{Gaia Collaboration} et~al., 2016, \mn@doi [\aap]
  {10.1051/0004-6361/201629272}, \href
  {https://ui.adsabs.harvard.edu/abs/2016A&A...595A...1G} {595, A1}

\bibitem[\protect\citeauthoryear{{Gaia Collaboration} et~al.,}{{Gaia
  Collaboration} et~al.}{2018}]{gaiacollaboration2018}
{Gaia Collaboration} et~al., 2018, \mn@doi [\aap]
  {10.1051/0004-6361/201833051}, \href
  {https://ui.adsabs.harvard.edu/abs/2018A&A...616A...1G} {616, A1}

\bibitem[\protect\citeauthoryear{{Gaia Collaboration} et~al.,}{{Gaia
  Collaboration} et~al.}{2022}]{GAIA3}
{Gaia Collaboration} et~al., 2022, arXiv e-prints, \href
  {https://ui.adsabs.harvard.edu/abs/2022arXiv220800211G} {p. arXiv:2208.00211}

\bibitem[\protect\citeauthoryear{{Gao}, {Xin}, {Liu}, {Zhang}  \& {Gao}}{{Gao}
  et~al.}{2016}]{gao16}
{Gao} Q.,  {Xin} Y.,  {Liu} J.-F.,  {Zhang} X.-B.,   {Gao} S.,  2016, \mn@doi
  [\apjs] {10.3847/0067-0049/224/2/37}, \href
  {https://ui.adsabs.harvard.edu/abs/2016ApJS..224...37G} {224, 37}

\bibitem[\protect\citeauthoryear{{Garrison} \& {Gray}}{{Garrison} \&
  {Gray}}{1994}]{gray94}
{Garrison} R.~F.,  {Gray} R.~O.,  1994, \mn@doi [\aj] {10.1086/116967}, \href
  {https://ui.adsabs.harvard.edu/abs/1994AJ....107.1556G} {107, 1556}

\bibitem[\protect\citeauthoryear{{Ghazaryan}, {Alecian}  \&
  {Hakobyan}}{{Ghazaryan} et~al.}{2018}]{ghazaryan2018}
{Ghazaryan} S.,  {Alecian} G.,   {Hakobyan} A.~A.,  2018, \mn@doi [\mnras]
  {10.1093/mnras/sty1912}, \href
  {https://ui.adsabs.harvard.edu/abs/2018MNRAS.480.2953G} {480, 2953}

\bibitem[\protect\citeauthoryear{{Ghazaryan}, {Alecian}  \&
  {Hakobyan}}{{Ghazaryan} et~al.}{2019}]{ghazaryan19}
{Ghazaryan} S.,  {Alecian} G.,   {Hakobyan} A.~A.,  2019, \mn@doi [\mnras]
  {10.1093/mnras/stz1678}, \href
  {https://ui.adsabs.harvard.edu/abs/2019MNRAS.487.5922G} {487, 5922}

\bibitem[\protect\citeauthoryear{{Gilliland} et~al.,}{{Gilliland}
  et~al.}{2010}]{SCdata}
{Gilliland} R.~L.,  et~al., 2010, \mn@doi [\apjl]
  {10.1088/2041-8205/713/2/L160}, \href
  {http://adsabs.harvard.edu/abs/2010ApJ...713L.160G} {713, L160}

\bibitem[\protect\citeauthoryear{{Gray} \& {Corbally}}{{Gray} \&
  {Corbally}}{2009}]{gray09}
{Gray} R.~O.,  {Corbally} C. J.,  2009, {Stellar Spectral Classification}

\bibitem[\protect\citeauthoryear{{Gray} \& {Garrison}}{{Gray} \&
  {Garrison}}{1987}]{gray87}
{Gray} R.~O.,  {Garrison} R.~F.,  1987, \mn@doi [\apjs] {10.1086/191237}, \href
  {http://adsabs.harvard.edu/abs/1987ApJS...65..581G} {65, 581}

\bibitem[\protect\citeauthoryear{{Gray} \& {Garrison}}{{Gray} \&
  {Garrison}}{1989a}]{gray89a}
{Gray} R.~O.,  {Garrison} R.~F.,  1989a, \mn@doi [\apjs] {10.1086/191315},
  \href {http://adsabs.harvard.edu/abs/1989ApJS...69..301G} {69, 301}

\bibitem[\protect\citeauthoryear{{Gray} \& {Garrison}}{{Gray} \&
  {Garrison}}{1989b}]{gray89b}
{Gray} R.~O.,  {Garrison} R.~F.,  1989b, \mn@doi [\apjs] {10.1086/191349},
  \href {http://adsabs.harvard.edu/abs/1989ApJS...70..623G} {70, 623}

\bibitem[\protect\citeauthoryear{{Gr{\"o}bel}, {H{\"u}mmerich}, {Paunzen}  \&
  {Bernhard}}{{Gr{\"o}bel} et~al.}{2017}]{groebel17}
{Gr{\"o}bel} R.,  {H{\"u}mmerich} S.,  {Paunzen} E.,   {Bernhard} K.,  2017,
  \mn@doi [\na] {10.1016/j.newast.2016.07.012}, \href
  {https://ui.adsabs.harvard.edu/abs/2017NewA...50..104G} {50, 104}

\bibitem[\protect\citeauthoryear{{Hamuy}, {Walker}, {Suntzeff}, {Gigoux},
  {Heathcote}  \& {Phillips}}{{Hamuy} et~al.}{1992}]{hamuy1992}
{Hamuy} M.,  {Walker} A.~R.,  {Suntzeff} N.~B.,  {Gigoux} P.,  {Heathcote}
  S.~R.,   {Phillips} M.~M.,  1992, \mn@doi [\pasp] {10.1086/133028}, \href
  {https://ui.adsabs.harvard.edu/abs/1992PASP..104..533H} {104, 533}

\bibitem[\protect\citeauthoryear{{Howell} et~al.,}{{Howell}
  et~al.}{2014}]{howell14}
{Howell} S.~B.,  et~al., 2014, \mn@doi [\pasp] {10.1086/676406}, \href
  {https://ui.adsabs.harvard.edu/abs/2014PASP..126..398H} {126, 398}

\bibitem[\protect\citeauthoryear{{H{\"u}mmerich} et~al.,}{{H{\"u}mmerich}
  et~al.}{2018}]{hummerich2018}
{H{\"u}mmerich} S.,  et~al., 2018, \mn@doi [\aap]
  {10.1051/0004-6361/201832938}, \href
  {https://ui.adsabs.harvard.edu/abs/2018A&A...619A..98H} {619, A98}

\bibitem[\protect\citeauthoryear{{H{\"u}mmerich}, {Paunzen}  \&
  {Bernhard}}{{H{\"u}mmerich} et~al.}{2020}]{huemmerich20}
{H{\"u}mmerich} S.,  {Paunzen} E.,   {Bernhard} K.,  2020, \mn@doi [\aap]
  {10.1051/0004-6361/202037750}, \href
  {https://ui.adsabs.harvard.edu/abs/2020A&A...640A..40H} {640, A40}

\bibitem[\protect\citeauthoryear{{Jamar}}{{Jamar}}{1977}]{jamar1977}
{Jamar} C.,  1977, \aap, \href
  {https://ui.adsabs.harvard.edu/abs/1977A&A....56..413J} {56, 413}

\bibitem[\protect\citeauthoryear{{Jenkins} et~al.,}{{Jenkins}
  et~al.}{2010}]{LCdata}
{Jenkins} J.~M.,  et~al., 2010, \mn@doi [\apjl] {10.1088/2041-8205/713/2/L120},
  \href {http://adsabs.harvard.edu/abs/2010ApJ...713L.120J} {713, L120}

\bibitem[\protect\citeauthoryear{{Koch} et~al.,}{{Koch} et~al.}{2010}]{koch10}
{Koch} D.~G.,  et~al., 2010, \mn@doi [\apjl] {10.1088/2041-8205/713/2/L79},
  \href {http://adsabs.harvard.edu/abs/2010ApJ...713L..79K} {713, L79}

\bibitem[\protect\citeauthoryear{{Krti{\v c}ka}, {Jan{\'{\i}}k}, {Markov{\'a}},
  {Mikul{\'a}{\v s}ek}, {Zverko}, {Prv{\'a}k}  \& {Skarka}}{{Krti{\v c}ka}
  et~al.}{2013}]{krticka13}
{Krti{\v c}ka} J.,  {Jan{\'{\i}}k} J.,  {Markov{\'a}} H.,  {Mikul{\'a}{\v s}ek}
  Z.,  {Zverko} J.,  {Prv{\'a}k} M.,   {Skarka} M.,  2013, \mn@doi [\aap]
  {10.1051/0004-6361/201221018}, \href
  {http://adsabs.harvard.edu/abs/2013A%26A...556A..18K} {556, A18}

\bibitem[\protect\citeauthoryear{{Krti{\v{c}}ka}, {Mikul{\'a}{\v{s}}ek},
  {L{\"u}ftinger}  \& {Jagelka}}{{Krti{\v{c}}ka} et~al.}{2015}]{krticka2015}
{Krti{\v{c}}ka} J.,  {Mikul{\'a}{\v{s}}ek} Z.,  {L{\"u}ftinger} T.,   {Jagelka}
  M.,  2015, \mn@doi [\aap] {10.1051/0004-6361/201425097}, \href
  {https://ui.adsabs.harvard.edu/abs/2015A&A...576A..82K} {576, A82}

\bibitem[\protect\citeauthoryear{{Krti{\v{c}}ka} et~al.,}{{Krti{\v{c}}ka}
  et~al.}{2019}]{krticka2019}
{Krti{\v{c}}ka} J.,  et~al., 2019, \mn@doi [\aap]
  {10.1051/0004-6361/201834937}, \href
  {https://ui.adsabs.harvard.edu/abs/2019A&A...625A..34K} {625, A34}

\bibitem[\protect\citeauthoryear{{Krti{\v{c}}ka}, {Mikul{\'a}{\v{s}}ek},
  {Prv{\'a}k}, {Niemczura}, {Leone}  \& {Wade}}{{Krti{\v{c}}ka}
  et~al.}{2020}]{krticka2020}
{Krti{\v{c}}ka} J.,  {Mikul{\'a}{\v{s}}ek} Z.,  {Prv{\'a}k} M.,  {Niemczura}
  E.,  {Leone} F.,   {Wade} G.,  2020, \mn@doi [\mnras]
  {10.1093/mnras/staa378}, \href
  {https://ui.adsabs.harvard.edu/abs/2020MNRAS.493.2140K} {493, 2140}

\bibitem[\protect\citeauthoryear{{Kurucz}}{{Kurucz}}{2005}]{kurucz2005}
{Kurucz} R.~L.,  2005, Memorie della Societa Astronomica Italiana Supplementi,
  \href {https://ui.adsabs.harvard.edu/abs/2005MSAIS...8...14K} {8, 14}

\bibitem[\protect\citeauthoryear{{Landstreet}}{{Landstreet}}{1982}]{landstreet82}
{Landstreet} J.~D.,  1982, \mn@doi [\apj] {10.1086/160114}, \href
  {https://ui.adsabs.harvard.edu/abs/1982ApJ...258..639L} {258, 639}

\bibitem[\protect\citeauthoryear{{Lanz}, {Artru}, {Le Dourneuf}  \&
  {Hubeny}}{{Lanz} et~al.}{1996}]{lanz96}
{Lanz} T.,  {Artru} M.~C.,  {Le Dourneuf} M.,   {Hubeny} I.,  1996, \aap, \href
  {https://ui.adsabs.harvard.edu/abs/1996A&A...309..218L} {309, 218}

\bibitem[\protect\citeauthoryear{{Leckrone}}{{Leckrone}}{1974}]{leckrone1974}
{Leckrone} D.~S.,  1974, \mn@doi [\apj] {10.1086/152879}, \href
  {https://ui.adsabs.harvard.edu/abs/1974ApJ...190..319L} {190, 319}

\bibitem[\protect\citeauthoryear{{Luo}, {Zhao}, {Zhao}  \& {et al.}}{{Luo}
  et~al.}{2018}]{DR4}
{Luo} A.~L.,  {Zhao} Y.~H.,  {Zhao} G.,   {et al.} 2018, VizieR Online Data
  Catalog, \href {https://ui.adsabs.harvard.edu/abs/2018yCat.5153....0L} {5153,
  0}

\bibitem[\protect\citeauthoryear{{Manfroid} \& {Mathys}}{{Manfroid} \&
  {Mathys}}{1986}]{manfroid86}
{Manfroid} J.,  {Mathys} G.,  1986, \aaps, \href
  {https://ui.adsabs.harvard.edu/abs/1986A&AS...64....9M} {64, 9}

\bibitem[\protect\citeauthoryear{{Martin} \& {GALEX Science Team}}{{Martin} \&
  {GALEX Science Team}}{2003}]{martin2003}
{Martin} C.,  {GALEX Science Team} 2003, in American Astronomical Society
  Meeting Abstracts. p. 96.01

\bibitem[\protect\citeauthoryear{{Masci} et~al.,}{{Masci} et~al.}{2019}]{ZTF2}
{Masci} F.~J.,  et~al., 2019, \mn@doi [\pasp] {10.1088/1538-3873/aae8ac}, \href
  {https://ui.adsabs.harvard.edu/abs/2019PASP..131a8003M} {131, 018003}

\bibitem[\protect\citeauthoryear{{Mathur} et~al.,}{{Mathur}
  et~al.}{2017}]{mathur2017}
{Mathur} S.,  et~al., 2017, \mn@doi [\apjs] {10.3847/1538-4365/229/2/30}, \href
  {https://ui.adsabs.harvard.edu/abs/2017ApJS..229...30M} {229, 30}

\bibitem[\protect\citeauthoryear{{Mikul{\'a}{\v{s}}ek}, {Zverko},
  {Krti{\v{c}}ka}, {Jan{\'\i}k}, {{\v{Z}}i{\v{z}}{\'n}ovsk{\'y}}  \&
  {Zejda}}{{Mikul{\'a}{\v{s}}ek} et~al.}{2007}]{mikulasek2007}
{Mikul{\'a}{\v{s}}ek} M.,  {Zverko} J.,  {Krti{\v{c}}ka} J.,  {Jan{\'\i}k} J.,
  {{\v{Z}}i{\v{z}}{\'n}ovsk{\'y}} J.,   {Zejda} M.,  2007, in {Romanyuk} I.~I.,
   {Kudryavtsev} D.~O.,  {Neizvestnaya} O.~M.,   {Shapoval} V.~M.,  eds,
  Physics of Magnetic Stars. pp 300--309

\bibitem[\protect\citeauthoryear{{Molnar}}{{Molnar}}{1973}]{molnar73}
{Molnar} M.~R.,  1973, \mn@doi [\apj] {10.1086/151892}, \href
  {http://adsabs.harvard.edu/abs/1973ApJ...179..527M} {179, 527}

\bibitem[\protect\citeauthoryear{{Molnar}}{{Molnar}}{1975}]{molnar75}
{Molnar} M.~R.,  1975, \mn@doi [\aj] {10.1086/111725}, \href
  {https://ui.adsabs.harvard.edu/abs/1975AJ.....80..137M} {80, 137}

\bibitem[\protect\citeauthoryear{{Molnar}, {Mallama}, {Holm}  \&
  {Soskey}}{{Molnar} et~al.}{1976}]{molnar1976}
{Molnar} M.~R.,  {Mallama} A.~D.,  {Holm} A.~V.,   {Soskey} D.~G.,  1976,
  \mn@doi [\apj] {10.1086/154703}, \href
  {https://ui.adsabs.harvard.edu/abs/1976ApJ...209..146M} {209, 146}

\bibitem[\protect\citeauthoryear{{Morrissey} et~al.,}{{Morrissey}
  et~al.}{2007}]{morrissey2007}
{Morrissey} P.,  et~al., 2007, \mn@doi [\apjs] {10.1086/520512}, \href
  {https://ui.adsabs.harvard.edu/abs/2007ApJS..173..682M} {173, 682}

\bibitem[\protect\citeauthoryear{{Nielsen}, {Gizon}, {Schunker}  \&
  {Karoff}}{{Nielsen} et~al.}{2013}]{nielsen13}
{Nielsen} M.~B.,  {Gizon} L.,  {Schunker} H.,   {Karoff} C.,  2013, \mn@doi
  [\aap] {10.1051/0004-6361/201321912}, \href
  {https://ui.adsabs.harvard.edu/abs/2013A&A...557L..10N} {557, L10}

\bibitem[\protect\citeauthoryear{{Olmedo}, {Lloyd}, {Mamajek}, {Ch{\'a}vez},
  {Bertone}, {Martin}  \& {Neill}}{{Olmedo} et~al.}{2015}]{olmedo2015}
{Olmedo} M.,  {Lloyd} J.,  {Mamajek} E.~E.,  {Ch{\'a}vez} M.,  {Bertone} E.,
  {Martin} D.~C.,   {Neill} J.~D.,  2015, \mn@doi [\apj]
  {10.1088/0004-637X/813/2/100}, \href
  {https://ui.adsabs.harvard.edu/abs/2015ApJ...813..100O} {813, 100}

\bibitem[\protect\citeauthoryear{{Osawa}}{{Osawa}}{1965}]{osawa65}
{Osawa} K.,  1965, Annals of the Tokyo Astronomical Observatory, \href
  {https://ui.adsabs.harvard.edu/abs/1965AnTok...9..121O} {9, 121}

\bibitem[\protect\citeauthoryear{{Paunzen}}{{Paunzen}}{2024}]{Paunzen2024}
{Paunzen} E.,  2024, \mn@doi [\aap] {10.1051/0004-6361/202348086}, \href
  {https://ui.adsabs.harvard.edu/abs/2024A&A...683L...7P} {683, L7}

\bibitem[\protect\citeauthoryear{{Paunzen} \& {Pri{\v{s}}egen}}{{Paunzen} \&
  {Pri{\v{s}}egen}}{2022}]{paunzen22}
{Paunzen} E.,  {Pri{\v{s}}egen} M.,  2022, \mn@doi [\aap]
  {10.1051/0004-6361/202244839}, \href
  {https://ui.adsabs.harvard.edu/abs/2022A&A...667L..10P} {667, L10}

\bibitem[\protect\citeauthoryear{{Paunzen}, {St{\"u}tz}  \&
  {Maitzen}}{{Paunzen} et~al.}{2005}]{paunzen05}
{Paunzen} E.,  {St{\"u}tz} C.,   {Maitzen} H.~M.,  2005, \mn@doi [\aap]
  {10.1051/0004-6361:20053001}, \href
  {https://ui.adsabs.harvard.edu/abs/2005A&A...441..631P} {441, 631}

\bibitem[\protect\citeauthoryear{{Preston}}{{Preston}}{1974}]{preston74}
{Preston} G.~W.,  1974, \mn@doi [\araa] {10.1146/annurev.aa.12.090174.001353},
  \href {http://adsabs.harvard.edu/abs/1974ARA%26A..12..257P} {12, 257}

\bibitem[\protect\citeauthoryear{{Prv{\'a}k}, {Li{\v{s}}ka}, {Krti{\v{c}}ka},
  {Mikul{\'a}{\v{s}}ek}  \& {L{\"u}ftinger}}{{Prv{\'a}k}
  et~al.}{2015}]{prvak2015}
{Prv{\'a}k} M.,  {Li{\v{s}}ka} J.,  {Krti{\v{c}}ka} J.,  {Mikul{\'a}{\v{s}}ek}
  Z.,   {L{\"u}ftinger} T.,  2015, \mn@doi [\aap]
  {10.1051/0004-6361/201526647}, \href
  {https://ui.adsabs.harvard.edu/abs/2015A&A...584A..17P} {584, A17}

\bibitem[\protect\citeauthoryear{{Renson} \& {Manfroid}}{{Renson} \&
  {Manfroid}}{2009}]{renson09}
{Renson} P.,  {Manfroid} J.,  2009, \mn@doi [\aap]
  {10.1051/0004-6361/200810788}, \href
  {https://ui.adsabs.harvard.edu/abs/2009A&A...498..961R} {498, 961}

\bibitem[\protect\citeauthoryear{{Samus}, {Kazarovets}, {Durlevich}, {Kireeva}
  \& {Pastukhova}}{{Samus} et~al.}{2017}]{GCVS}
{Samus} N.~N.,  {Kazarovets} E.~V.,  {Durlevich} O.~V.,  {Kireeva} N.~N.,
  {Pastukhova} E.~N.,  2017, \mn@doi [Astronomy Reports]
  {10.1134/S1063772917010085}, \href
  {http://adsabs.harvard.edu/abs/2017ARep...61...80S} {61, 80}

\bibitem[\protect\citeauthoryear{{Shulyak}, {Krti{\v{c}}ka},
  {Mikul{\'a}{\v{s}}ek}, {Kochukhov}  \& {L{\"u}ftinger}}{{Shulyak}
  et~al.}{2010}]{shulyak10}
{Shulyak} D.,  {Krti{\v{c}}ka} J.,  {Mikul{\'a}{\v{s}}ek} Z.,  {Kochukhov} O.,
   {L{\"u}ftinger} T.,  2010, \mn@doi [\aap] {10.1051/0004-6361/201015094},
  \href {https://ui.adsabs.harvard.edu/abs/2010A&A...524A..66S} {524, A66}

\bibitem[\protect\citeauthoryear{{Siegmund} et~al.,}{{Siegmund}
  et~al.}{2004}]{siegmund2004}
{Siegmund} O. H.~W.,  et~al., 2004, in {Hasinger} G.,  {Turner} M. J.~L.,  eds,
   Society of Photo-Optical Instrumentation Engineers (SPIE) Conference Series
  Vol. 5488, UV and Gamma-Ray Space Telescope Systems. pp 13--24,
  \mn@doi{10.1117/12.561488}

\bibitem[\protect\citeauthoryear{{Smalley}}{{Smalley}}{2004}]{smalley2004}
{Smalley} B.,  2004, in {Zverko} J.,  {Ziznovsky} J.,  {Adelman} S.~J.,
  {Weiss} W.~W.,  eds,  IAU Symposium Vol. 224, The A-Star Puzzle. pp 131--138
  (\mn@eprint {arXiv} {astro-ph/0408222}), \mn@doi{10.1017/S1743921304004478}

\bibitem[\protect\citeauthoryear{{Sokolov}}{{Sokolov}}{2000}]{sokolov2000}
{Sokolov} N.~A.,  2000, \aap, \href
  {https://ui.adsabs.harvard.edu/abs/2000A&A...353..707S} {353, 707}

\bibitem[\protect\citeauthoryear{{Sokolov}}{{Sokolov}}{2011}]{sokolov2011}
{Sokolov} N.~A.,  2011, in Magnetic Stars. pp 390--398 (\mn@eprint {arXiv}
  {1104.1547}), \mn@doi{10.48550/arXiv.1104.1547}

\bibitem[\protect\citeauthoryear{{Sokolov}}{{Sokolov}}{2012}]{sokolov2012}
{Sokolov} N.~A.,  2012, \mn@doi [\mnras] {10.1111/j.1365-2966.2011.19926.x},
  \href {https://ui.adsabs.harvard.edu/abs/2012MNRAS.426.2819S} {426, 2819}

\bibitem[\protect\citeauthoryear{{Stassun} et~al.,}{{Stassun}
  et~al.}{2019}]{2019AJ....158..138S}
{Stassun} K.~G.,  et~al., 2019, \mn@doi [\aj] {10.3847/1538-3881/ab3467}, \href
  {https://ui.adsabs.harvard.edu/abs/2019AJ....158..138S} {158, 138}

\bibitem[\protect\citeauthoryear{{Stibbs}}{{Stibbs}}{1950}]{stibbs50}
{Stibbs} D.~W.~N.,  1950, \mn@doi [\mnras] {10.1093/mnras/110.4.395}, \href
  {http://adsabs.harvard.edu/abs/1950MNRAS.110..395S} {110, 395}

\bibitem[\protect\citeauthoryear{{Thompson}, {Fraquelli}, {Van Cleve}  \&
  {Caldwell}}{{Thompson} et~al.}{2016}]{thompson2016}
{Thompson} S.~E.,  {Fraquelli} D.,  {Van Cleve} J.~E.,   {Caldwell} D.~A.,
  2016, {Kepler Archive Manual}, Kepler Science Document KDMC-10008-006, id. 9.
  Edited by Faith Abney, Dwight Sanderfer, Michael R. Haas, and Steve B. Howell

\bibitem[\protect\citeauthoryear{{Tody}}{{Tody}}{1986}]{tody1986}
{Tody} D.,  1986, in {Crawford} D.~L.,  ed.,  Society of Photo-Optical
  Instrumentation Engineers (SPIE) Conference Series Vol. 627, Instrumentation
  in astronomy VI. p.~733, \mn@doi{10.1117/12.968154}

\bibitem[\protect\citeauthoryear{{Tody}}{{Tody}}{1993}]{tody1993}
{Tody} D.,  1993, in {Hanisch} R.~J.,  {Brissenden} R.~J.~V.,   {Barnes} J.,
  eds,  Astronomical Society of the Pacific Conference Series Vol. 52,
  Astronomical Data Analysis Software and Systems II. p.~173

\bibitem[\protect\citeauthoryear{{Wenger} et~al.,}{{Wenger}
  et~al.}{2000}]{SIMBAD}
{Wenger} M.,  et~al., 2000, \mn@doi [\aaps] {10.1051/aas:2000332}, \href
  {https://ui.adsabs.harvard.edu/abs/2000A&AS..143....9W} {143, 9}

\bibitem[\protect\citeauthoryear{{Wolff} \& {Wolff}}{{Wolff} \&
  {Wolff}}{1971}]{wolff71}
{Wolff} S.~C.,  {Wolff} R.~J.,  1971, \mn@doi [\aj] {10.1086/111138}, \href
  {http://adsabs.harvard.edu/abs/1971AJ.....76..422W} {76, 422}

\bibitem[\protect\citeauthoryear{{Zhang}, {Green}  \& {Rix}}{{Zhang}
  et~al.}{2023}]{2023MNRAS.524.1855Z}
{Zhang} X.,  {Green} G.~M.,   {Rix} H.-W.,  2023, \mn@doi [\mnras]
  {10.1093/mnras/stad1941}, \href
  {https://ui.adsabs.harvard.edu/abs/2023MNRAS.524.1855Z} {524, 1855}

\bibitem[\protect\citeauthoryear{Zhao, Zhao, Chu, Jing  \& Deng}{Zhao
  et~al.}{2012}]{lamost1}
Zhao G.,  Zhao Y.-H.,  Chu Y.-Q.,  Jing Y.-P.,   Deng L.-C.,  2012, Research in
  Astronomy and Astrophysics, 12, 723

\makeatother
\end{thebibliography}



\appendix

\section{The phase diagrams}
\label{sec:appendixa}

In this Section, we present the phase diagrams of all the 28 stars of our sample.

\begin{figure*}
\centering
\begin{tabular}{ccc}
\resizebox{0.30\textwidth}{!}{\includegraphics{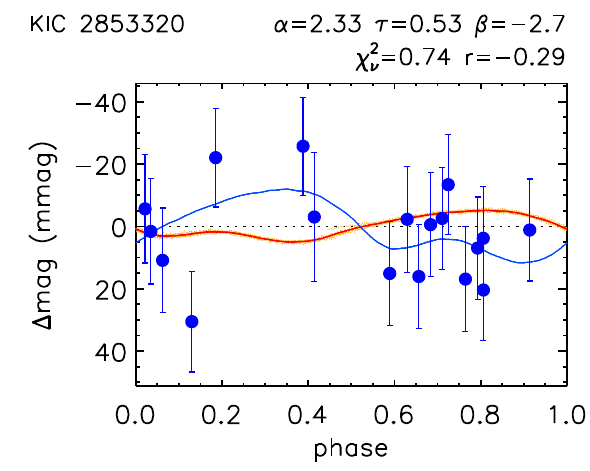}} &
\resizebox{0.30\textwidth}{!}{\includegraphics{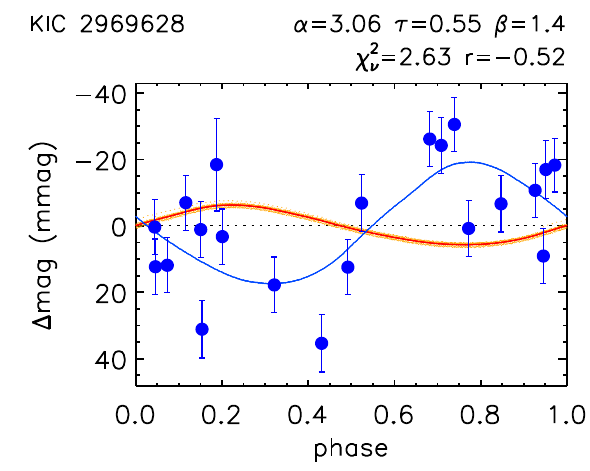}} &
\resizebox{0.30\textwidth}{!}{\includegraphics{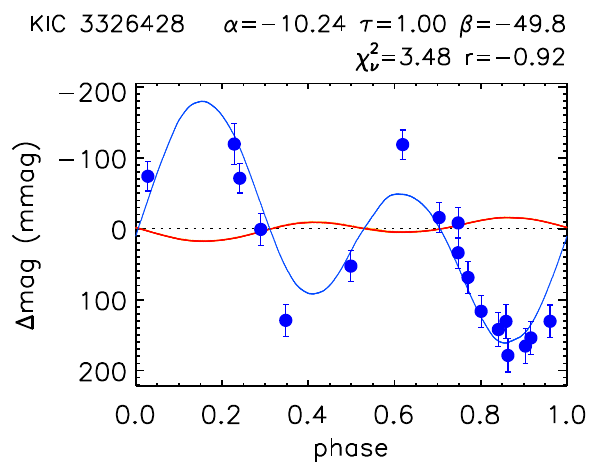}} \\

\resizebox{0.30\textwidth}{!}{\includegraphics{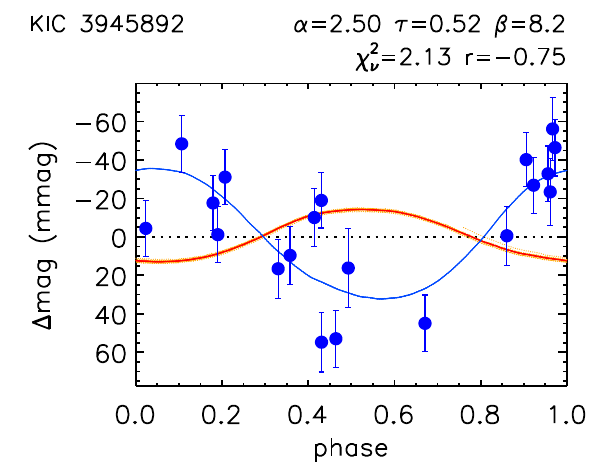}} &
\resizebox{0.30\textwidth}{!}{\includegraphics{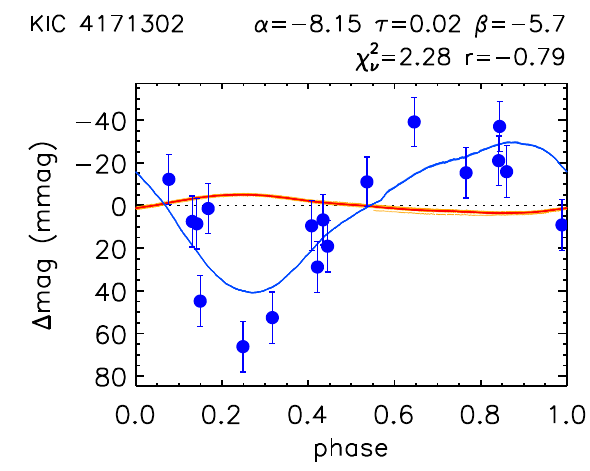}} &
\resizebox{0.30\textwidth}{!}{\includegraphics{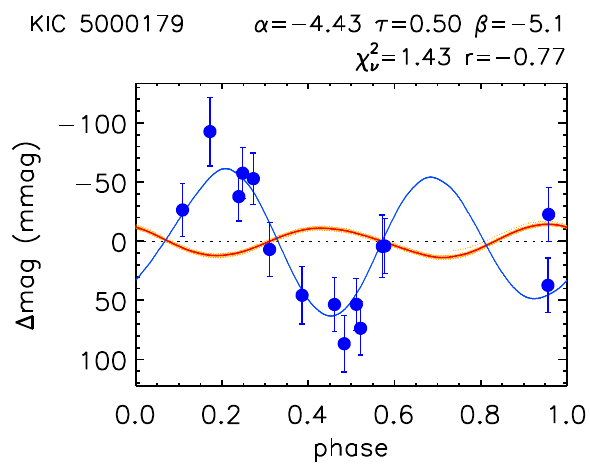}} \\

\resizebox{0.30\textwidth}{!}{\includegraphics{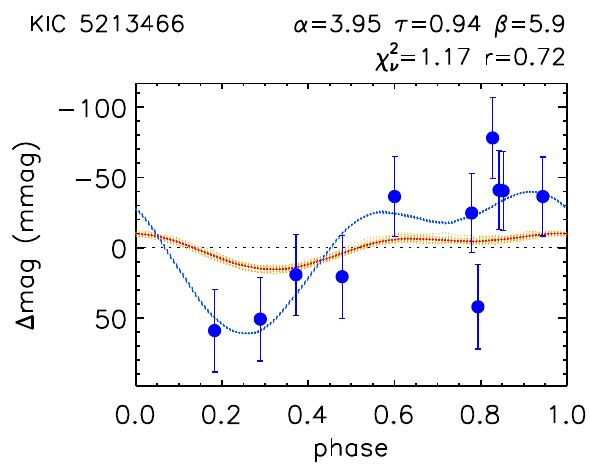}} &
\resizebox{0.30\textwidth}{!}{\includegraphics{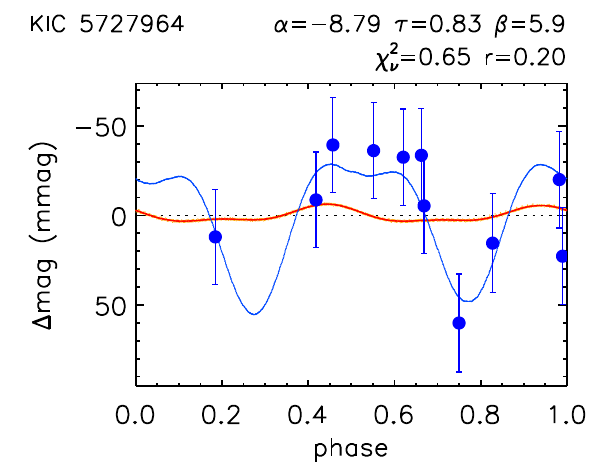}} &
\resizebox{0.30\textwidth}{!}{\includegraphics{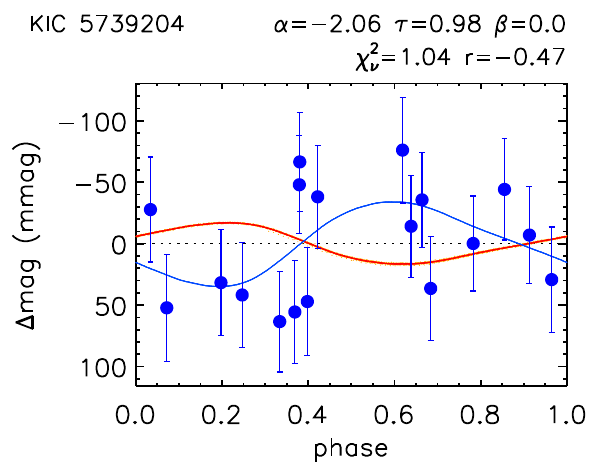}} \\

\resizebox{0.30\textwidth}{!}{\includegraphics{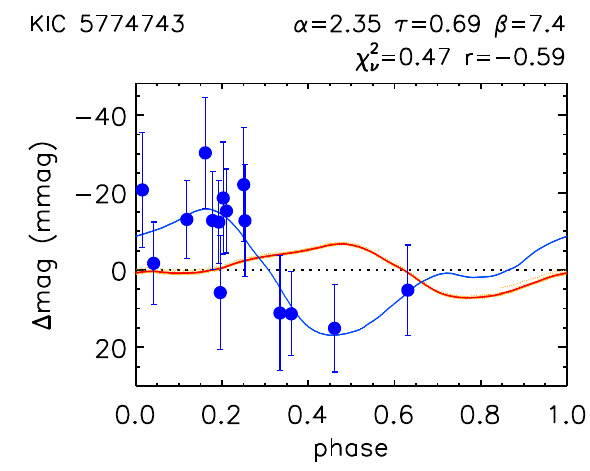}} &
\resizebox{0.30\textwidth}{!}{\includegraphics{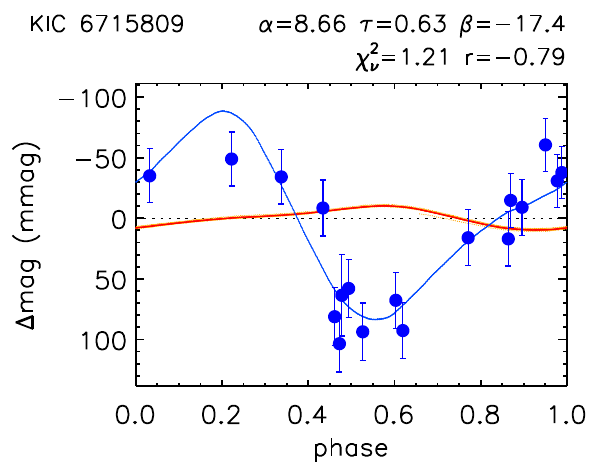}} &
\resizebox{0.30\textwidth}{!}{\includegraphics{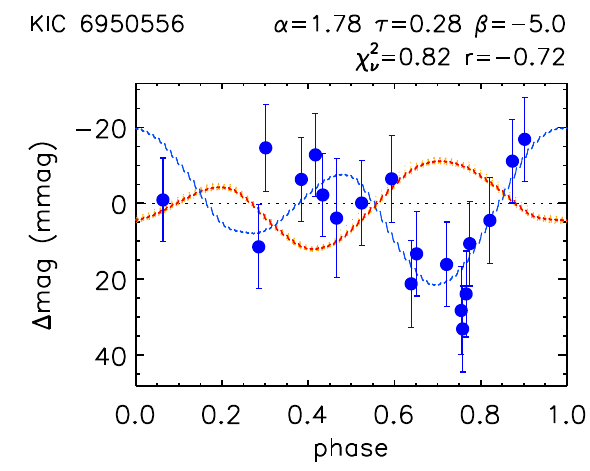}} \\

\resizebox{0.30\textwidth}{!}{\includegraphics{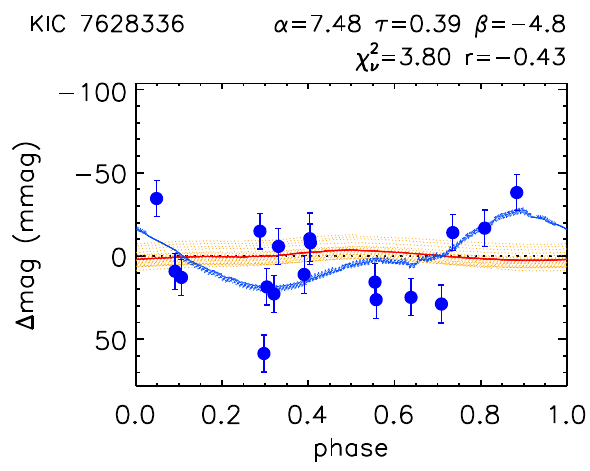}} &
\resizebox{0.30\textwidth}{!}{\includegraphics{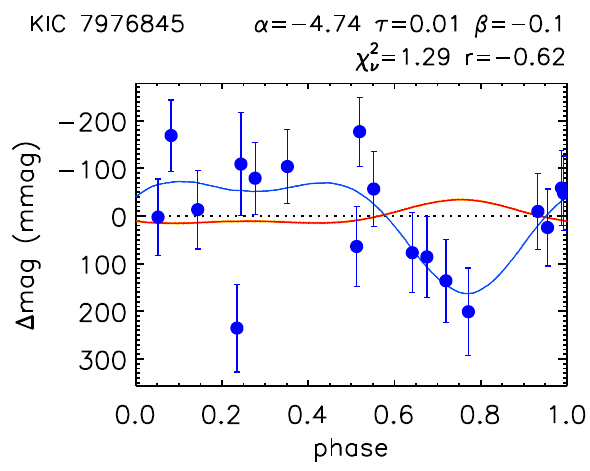}} &
\resizebox{0.30\textwidth}{!}{\includegraphics{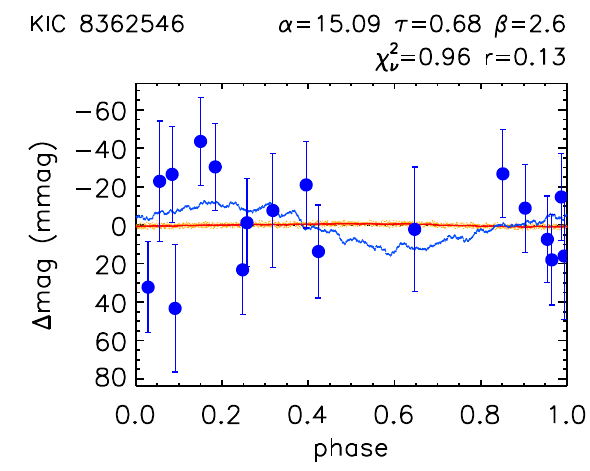}} \\

\end{tabular}
\caption{Phase diagrams for the stars of our sample. Orange dots are the $\sigma$-clipped {\em Kepler} data, while the smoothed curve is shown in red. The blue curve is the best fit of the {\em GALEX} points (large blue dots), obtained from the genetic algorithm, as explained in Sect.~\ref{sec:aga}. On top of each panel, the KIC ID and some of the results of Table~\ref{tab:agaresults} are reported.}
\label{fig:appendixphasediagrams}
\end{figure*}
 
\begin{figure*}
\centering
\begin{tabular}{ccc}
\resizebox{0.30\textwidth}{!}{\includegraphics{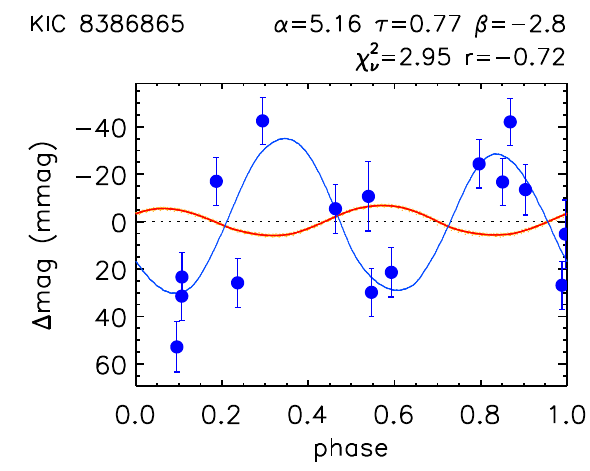}} &
\resizebox{0.30\textwidth}{!}{\includegraphics{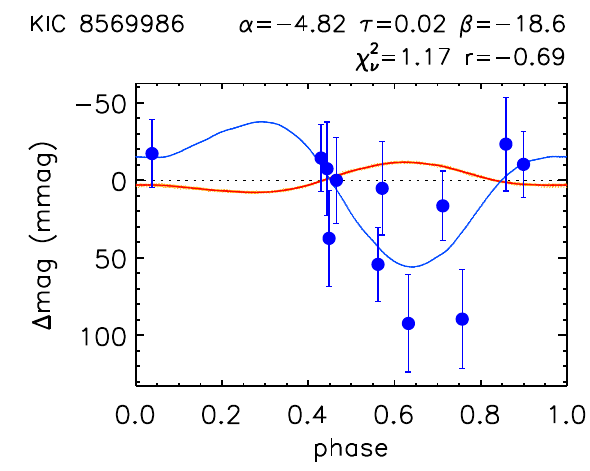}} &
\resizebox{0.30\textwidth}{!}{\includegraphics{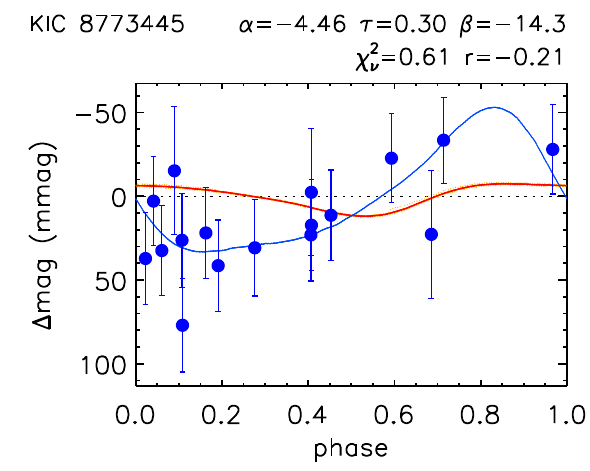}} \\

\resizebox{0.30\textwidth}{!}{\includegraphics{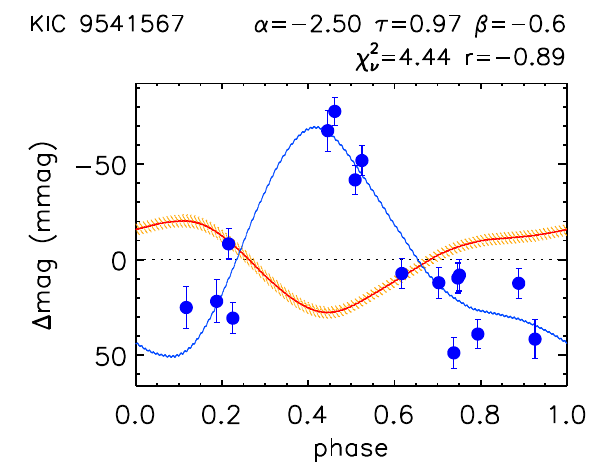}} &
\resizebox{0.30\textwidth}{!}{\includegraphics{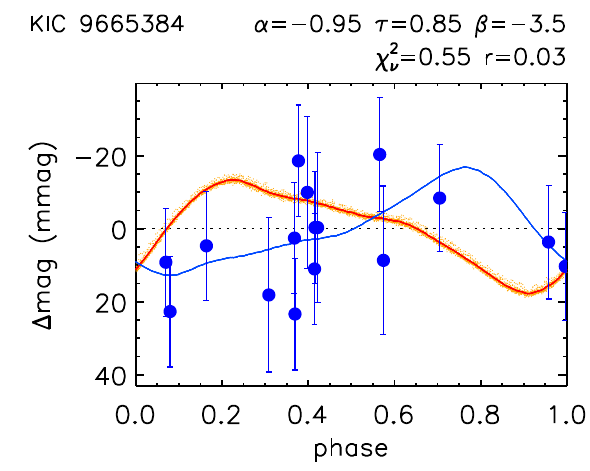}} &
\resizebox{0.30\textwidth}{!}{\includegraphics{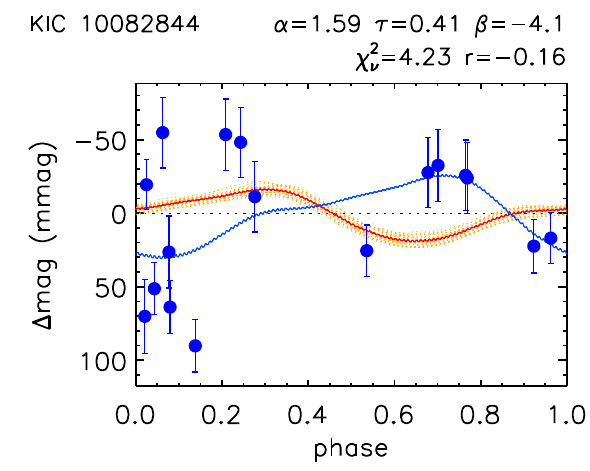}} \\

\resizebox{0.30\textwidth}{!}{\includegraphics{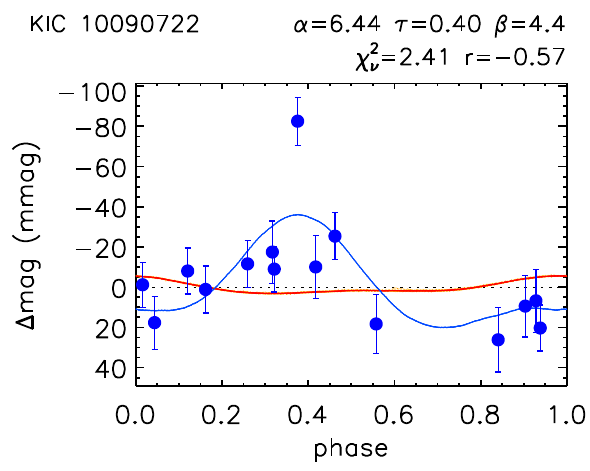}} &
\resizebox{0.30\textwidth}{!}{\includegraphics{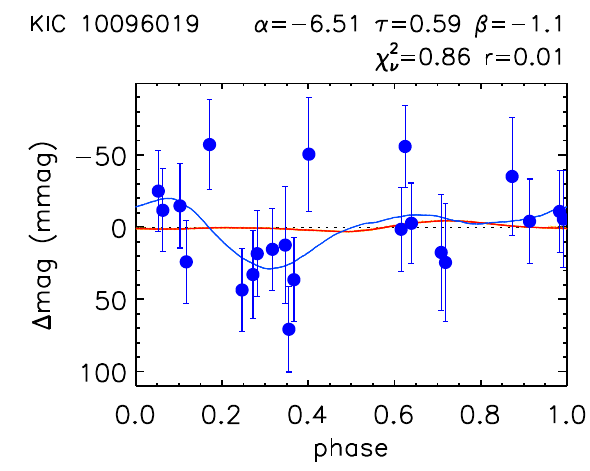}} &
\resizebox{0.30\textwidth}{!}{\includegraphics{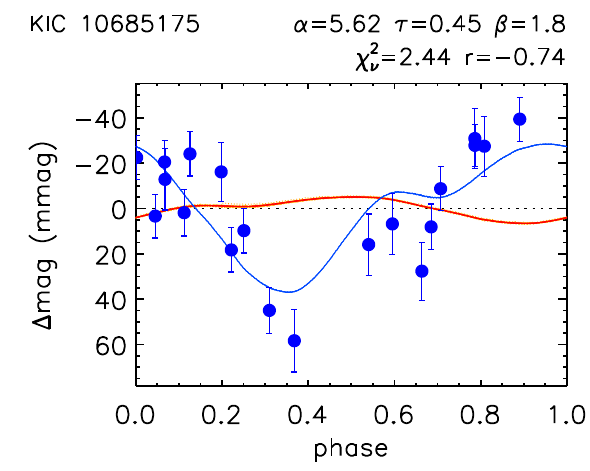}} \\

\resizebox{0.30\textwidth}{!}{\includegraphics{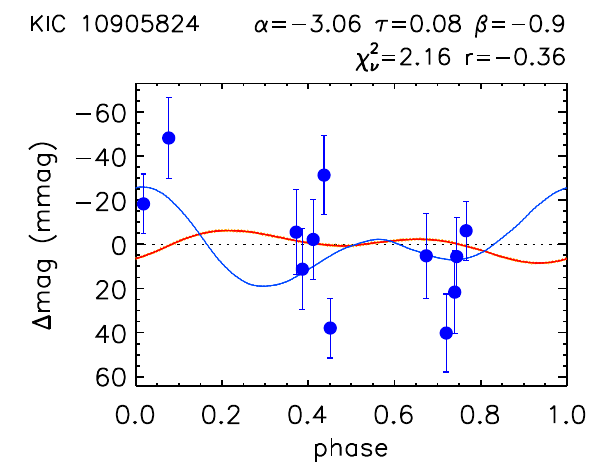}} &
\resizebox{0.30\textwidth}{!}{\includegraphics{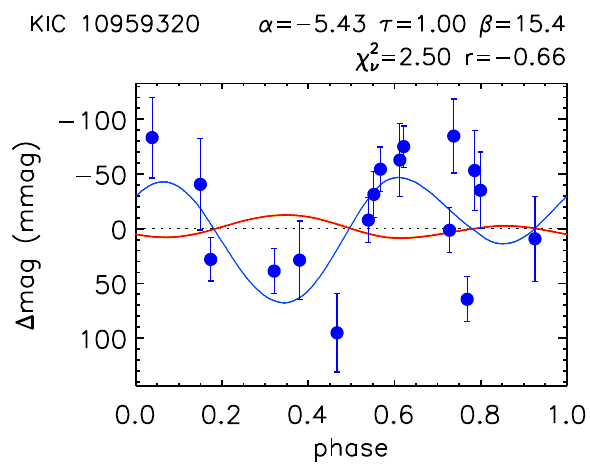}} &
\resizebox{0.30\textwidth}{!}{\includegraphics{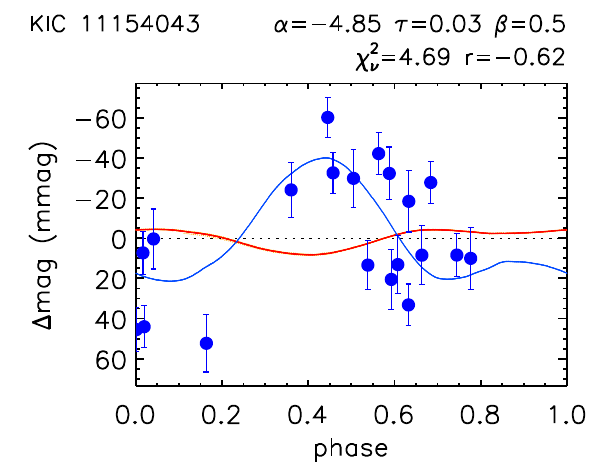}} \\

\resizebox{0.30\textwidth}{!}{\includegraphics{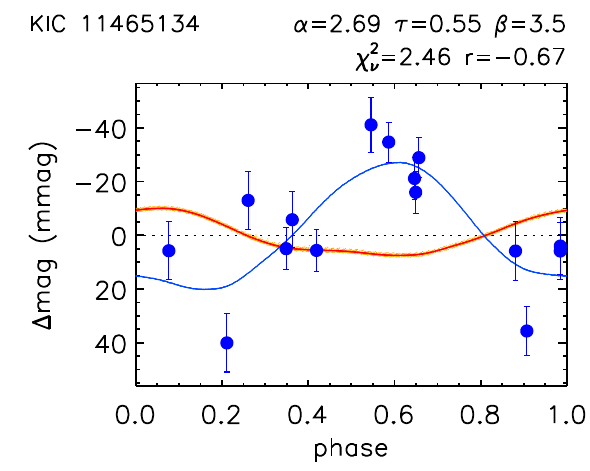}} & & \\

\end{tabular}
\caption{Same as in Fig.~\ref{fig:appendixphasediagrams}, but for the remaining stars of the sample.}
\end{figure*}


\bsp	
\label{lastpage}
\end{document}